\documentclass[a4paper]{article}
\usepackage{graphics,epsfig}
\usepackage{graphicx}
\title {Intrinsic Harmonic Spaces: A Solution to the \\ Ancient Problem of Perfect Tuning}
\author {Diederik Aerts}
\date {}
\newcommand{\sep}{\hskip -2pt \makebox[11pt]{$\bigcirc$ \hskip -11.5 pt
\raise 0.1 pt \hbox{$\wedge$}}}

\newcommand{\smallsep}{\footnotesize\hskip -2pt \makebox[11pt]{$\bigcirc$
\hskip -10 pt \raise 0.1 pt \hbox{$\wedge$}}}

\font\smallroman=cmr10 at 7pt
\font\mediumroman=cmr10 at 8pt
\def\be{\begin{equation}}
\def\ee{\end{equation}}
\def\bea{\begin{eqnarray}}
\def\eea{\end{eqnarray}}
\usepackage{indentfirst}
\usepackage[psamsfonts]{amssymb}
\usepackage{longtable,lscape}

\newcommand{\integ}{{\mathbb Z}}

\begin{document}

\maketitle
\centerline{Center Leo Apostel (CLEA)}
\centerline{Vrije Universiteit Brussel,}
\centerline{Krijgskundestraat 33, 1160 Brussels.}
\centerline{e-mail: diraerts@vub.ac.be}

\begin{abstract}
\noindent 
In this article we solve this ancient problem of perfect tuning in all keys and present a system were all harmonies are conserved at once. It will become clear, when we expose our solution, why this solution could not be found in the way in which earlier on musicians and scientist have been approaching the problem. We follow indeed a different approach. We first construct a mathematical representation of the complete harmony by means of a vector space, where the different tones are represented in complete harmonic way for all keys at once. One of the essential differences with earlier systems is that tones will no longer be ordered within an octave, and we find the octave-like ordering back as a projection of our system. But it is exactly by this projection procedure that the possibility to create a harmonic system for all keys at once is lost. So we see why the old way of ordering tones within an octave could not lead to a solution of the problem. We indicate in which way a real musical instrument could be built that realizes our harmonic scheme. Because tones are no longer ordered within an octave such a musical instrument will be rather unconventional. It is however a physically realizable musical instrument, at least for the Pythagorean harmony. We also indicate how perfect harmonies of every dimension could be realized by computers.
\end{abstract}

\section{Introduction}
\noindent
In earlier times instrument tuning was anything but a standard art \cite{owen,sullivan,isacoff}. Already in the days of Pythagoras it was recognized that there are problems with creating a perfectly tuned scale \cite{abraham,grout}.  Over the centuries there have been many attempts to create tuning schemes that preserve the harmony of perfectly tuned intervals. The main approach was to try to minimize the errors that naturally occur when trying to solve the problem of harmony. That is the reason that during the history of music many different tuning systems have been proposed. They differ mostly with respect to what type of harmonies they try to conserve at the cost of loosing other ones. But none of them could produce a full harmony for all keys, and it has been generally accepted that it is impossible to do so \cite{owen,abraham,grout}.

The tuning scheme that is now generally adopted in our western society is called the `equal temperament' system. But many tuning systems have been common place during the history of western music, and in non-western music even nowadays alternative tuning systems are used. In this article we will study the general problem of harmony and clarify how the struggle with this problem has lead to all the different tuning systems. Although the solution that we present in this article, the construction of intrinsic harmonic spaces, does not really fit into the historical scheme, {\it i.e.} it is not a new tuning scheme, to be able to explain our solution as compared to traditional musical theory, we will have to use the concepts that have been introduced over the years. Let us therefore start by introducing these concepts and also their standard notations.

Notes, tones and tuning systems are the subject of our reflection. The collection of notes consists of the
traditional twelve half notes in the different octaves. A tone can be uniquely identified by giving its frequency. Frequency is measured in a unity that is called Hertz, expressing the number of cycles per second. Traditionally the tone $a$ of the central octave of a keyboard is tuned to 440 Hertz. A tuning systems defines a way to make correspond to each note a specific tone. 

\subsection{The Notes}
\noindent
We will use the
common naming for the twelve half notes that are used in an octave in western music with $c$ as base note:
$c$, $d_b$, $d$, $e_b$, $e$, $f$, $f^\#$, $g$, $a_b$, $a$, $b_b$, and $b$.
This are the notes of one octave, that we will consider as the basic octave, and we will refer to the notes of this
octave as the basic notes. The notes of higher octaves to this basic octave we denote by using a positive whole number as
subscript, the number indicating the different higher octaves in increasing order. This means for example that $a_1$ is
the note $a$ of the first octave higher than the basic octave and $b_3$ is the note $b$ of the third octave higher than
the basic octave. The notes of the lower octaves to this basic octave we denote by using a negative whole number as
subscript, the absolute value of the number indicating the different lower octaves in decreasing order. This means for
example that $c_{-2}$ is the note $c$ of the second octave lower to the basic octave. Taking the logic of this notation
to its limit, we should notate a note of the basic octave, for example the note $a$, as $a_0$, and sometimes, if this
comes out better for the formulas we will do so. Sometimes, when we want to make a statement about a general note, we
will indicate this note by a variable, and use the variables $q, r$ and $s$ to indicate notes. We have introduced all
the rules of notation to define now the set of all notes, that we will denote by ${\cal N}$. The set of all notes that we consider is given by
\be
{\cal N} = \{q_k\ \vert\ q \in \{c, d_b, d, e_b, e, f, g_b, g, a_b, a, b_b, b\}, k \in \integ \} 
\ee
where $\{c, d_b, d, e_b, e, f, g_b, g, a_b, a, b_b, b\}$ is the set of notes in the basic octave, and $\integ$ is the
set of integers, positive as well as negative.

\subsection{The Tones and the Tuning System} \label{sec:tones}
\noindent
Tones will be expressed basically in Hertz, and as we mentioned already, the tone $a$ correspondng to the note $a$ is
usually tuned at 440 Hertz. We will indicate tones by means of the letters $\{u, v, w, \ldots\}$.

The harmonic content of two
different tones is directly related to the ratio of their frequencies being a simple number. If we consider two
tones that have the distance of an entire octave between them, for example the tones corresponding to the notes
$c$ and
$c_1$, then the frequency of the highest tone, the one corresponding to the note $c_1$, is exactly the double of the
frequency of the lower tone, the one corresponding to the note
$c$. This is the case for all tuning systems. This means that from now on, independent of the adopted tuning
system, the relation between $q_x$ where $q$ is one of the half notes of octave $x$, and $q_{x+1}$, which is the same
notes of octave $x+1$, is such that that the frequency of the tone corresponding to the note $q_{x+1}$ is the double of
the frequency of the tone corresponding to the note
$q_x$. That is the reason that
$c$ and $c_1$ sound very harmonically together; in this case we even hear it as the same tone, but lower and
higher in pitch.

If we consider the notes $c$ and $g_1$, then, for a perfect tuning system, the frequency corresponding to $g_1$
is the triple of the frequency corresponding to $c$. This is not the case for the `equal temperament' tuning
system, the one that is now adopted in our western society. That is the reason that in most ancient tuning systems also
these two notes,
$c$ and $g_1$ sound very harmonically together. These notes sound less harmonically together when played on
actual instruments, because of equal temperament that is now universally adopted.

Let us explain what `equal temperament' is. For reasons that we will come back to in section \ref{section04},
it was decided to have twelve half notes in one octave, and to tune these half notes in such a way that the ratio
between the frequencies corresponding to two consecutives half notes is the same for all couples of consecutives half notes.  Let us introduce first some more notations. When we talk of an arbitrary note, we will denote this
note by using one of the letters $q, r,$ or $s$. Octaves we will denote by $x$ running over
the set of whole numbers $\{\ldots, -3, -2, -1, 0, 1, 2, 3, \ldots\}$, where
$0$ corresponds to what we decide to be our basic octave. When we want to specify for a note $q$ that it is a note of the
octave $x$ we write $q_x$. We denote arbitrary tones by the letters $u, v, w, \ldots$. The different tuning systems
that we will talk about, we will indicate by capital letters, for example the equal temperament tuning system
is denoted by $E$, the perfect harmonic system by $H$, and the Pythagorean tuning system by $P$. If we want to denote
the tone
$u$ that within a tuning system $K$ corresponds to the note $q$ we write $u = u(K,q)$. Hence $u(E,q)$ is the tone that
corresponds to the note
$q$ in the equal temperament system, and $u(H,q)$ is the tone that corresponds to note $q$ in the perfect harmonic
system, while
$u(P,q)$ is the tone that corresponds to the note
$q$ in the Pythagorean tuning system. When the frequency of the
tone
$v$ corresponding to note
$r$ within the tuning system $K$ is
$k$ times the frequency of the tone $u$ corresponding to the note $q$ within the tuning system $L$ we denote
\be
v(K,r)/u(L,q) = k \label{v/u}
\ee 
This means that we can write now for the notes $c$ and $c_1$ the following
\be
u(E,c_1)/u(E,c) = 2
\ee
and also
\be
u(H,c_1)/u(H,c) = u(P,c_1)/u(P,c) = 2
\ee
Let us calculate the frequency ratio between two consecutives half notes of the same octave in the case
of equal temperament. Hence $c, d_b, d, e_b, e, f, g_b, g, a_b, a, b_b, b, c_1$ is the set of twelve half notes that we
consider. Equal temperament means that
\bea
&&k = u(E,d_b)/u(E,c) = u(E,d)/u(E,d_b) = u(E,e_b)/u(E,d) = u(E,e)/u(E,e_b) \\
&&= u(E,f)/u(E,e) = u(E,g_b)/u(E,f) = u(E,g)/u(E,g_b) = u(E,a_b)/u(E,g) \\
&&= u(E,a)/u(E,a_b) = u(E,b_b)/u(E,a) = u(E,b)/u(E,b_b) = u(E,c_1)/u(E,b)
\eea
We also have
\bea
2 = u(E,c_1)/u(E,c) &=& u(E,d_b)/u(E,c) \cdot u(E,d)/u(E,d_b) \cdot u(E,e_b)/u(E,d) \\
&\cdot& u(E,e)/u(E,e_b) \cdot u(E,f)/u(E,e) \cdot u(E,g_b)/u(E,f) \\
&\cdot& u(E,g)/u(E,g_b) \cdot  u(E,a_b)/u(E,g) \cdot u(E,a)/u(E,a_b) \\
&\cdot& u(E,b_b)/u(E,a) \cdot u(E,b)/u(E,b_b) \cdot u(E,c_1)/u(E,b) \\
&=& k^{12}
\eea
and as a consequence
\be
k^{12} = 2 \Leftrightarrow k = {\root 12 \of 2}
\ee
If we calculate $\root 12 \of 2$ we find $k = 1.059463$. What is more important to remark however is that $\root 12 \of
2$ is an irrational number, which means that there does not exist a fraction that equals this number. It also means that
the complete equal tempering deviates in principle completely from the base of harmony itself. It only allows a rational
number (fraction of whole numbers) for the relation of the frequencies of basic tones of the octaves, but no rational
numbers for all of the other relations of frequencies. What is it that made adopt standard western music this completely
non harmonic tuning system? There is a long history of trying out different tuning systems involved and a long struggle
with harmony. We will count partly this story in the following of this article while we introduce our solution to the
problem, the intrinsic harmonic spaces.

\subsection{Harmonic Possibilities} \label{sec:harmonicpossibilities}
\noindent
Before the completely equal tempered tuning system was adopted, many other much more harmonic systems have been in use.
Let us analyse the possible harmonic solutions to tuning. The fifth represents, after the octave, the next
building block of harmony. If we consider the note $g_1$, hence the $sol$ of the octave which is one higher than the
basic octave, in a perfect harmonic tuning system, the frequency corresponding to this note $g_1$ is three times
the frequency of the basic note $c$. Hence
\be
u(PYT,g_1)/u(PYT,c) = 3
\ee
This gives us the information to calculate the ratio of the fifth itself, namely the ratio $u(PYT,g)/u(PYT,c)$. Indeed, we
have
\be
u(H,g_1)/u(H,g) = 2
\ee
which shows that
\be
u(H,g)/u(H,c) = u(H,g)/u(H,g_1) \cdot u(H,g_1)/u(H,c) = 1/2 \cdot 3 = 3/2
\ee
So the frequency relation between a tone and its fifth is $3/2$ in a harmonically tuned system. We also know that
$c$ is the fifth for
$f_{-1}$, which means that
\be
u(H,f_{-1})/u(H,c) = 2/3
\ee
and hence
\be
u(H,f)/u(H,c) = 4/3
\ee
The relation between $c$ and $f$ is called a fourth in music theory. These results are presented in Table \ref{tablebasictunes}.

\begin{table}[h]
\caption{The ratio's of basic, fourth, fifth and octave in the case of
perfect and tempered tuning}
\begin{center}
\begin{tabular}{llllll} \hline 
\multicolumn{2}{l}{} & \multicolumn{2}{l}{\bf Harmonic} & \multicolumn{2}{l}{\bf Tempered} \\
\hline   
{\it Tone Ratio} & {\it Example} & {\it Freq. Fraction}  & {\it Cent} & {\it Freq. Fraction} &
{\it Cent} \\ \hline Basic & $u(c)/u(c)$ & 1 & 0.000 & 1 & 0.000 \\ 
Fourth & $u(f)/u(c)$ & 4/3 = 1.3333 & 498.044 & $(\root 12 \of {2})^5$ = 1.3348 & 500.000\\ 
Fifth & $u(g)/u(c)$ & 3/2 = 1.5 & 701.955 & $(\root 12 \of {2})^7$ = 1.4983 & 700.000\\ 
Octave & $u(c_1)/u(c)$ & 2 & 1200.000 & 2 & 1200.000\\ \hline
\end{tabular}
\label{tablebasictunes}
\end{center}
\end{table}
\noindent
In the fourth column of Table \ref{tablebasictunes} we have represented the ratio's in cent, which is the commonly used measure
for pitch differences in traditional music theory. It is an algorithmic measure, where ratio $1$ corresponds to $0.000$
cent and ratio $2$ corresponds to $1200.000$ cent. The following formula can be used to calculate the difference in pitch
given the ratio's of the two tones. If $u$ and $v$ are the two tones, and hence their ratio is given by $v/u$, then
\be
{1200 \over \log2} \cdot \log{v/u}
\ee
gives the difference in pitch of tone $u$ and $v$ measured in cent. The fraction
$1200/\log2$ is a renormalization factor, otherwise the octave would
correspond to the interval $[0, \log2]$ instead of the interval $[0, 1200]$. Remark that the cent measure is introduced
with reference to the completely equal tempered tuned system. Indeed, suppose that we consider two tones $u$ and $v$ that
correspond to two consecutive half notes in a completely equal tempered tuning system. Then $v/u = {\root 12 \of 2}$.
Let us calculate the difference in cent of these two tones. We have 
\bea
{1200 \over \log2} \cdot \log{v/u} &=& {1200 \over \log2} \cdot \log{\root 12 \of 2} \\
&=& {1200 \over \log2} \cdot \log2/12 \\
&=& 100
\eea
This means that in the case of a completely equal tempered system, the 1200 cent of an octave are equally divided into
12 times 100 cent for each half tone step. Remark that indeed in Table \ref{tablebasictunes}, the harmonic system gives simple
ratio's for the different tone ratio's, and complicated cent measures (see Table \ref{tablebasictunes} {\bf Harmonic}), while the
tempered system gives simple cent measures, and complicated ratio's (see Table \ref{tablebasictunes} {\bf Tempered}).

Let us proceed analyzing harmony. Starting with $c$ as basic tone, Table \ref{tablebasictunes} gives us the perfect tuning
ratio's corresponding to
$f$ (fourth), $g$ (fifth) and $c_1$ (octave). What about the third, or $e$ in
case we take $c$ as basic tone? Here different possible perfect tuning solutions appear. The
oldest proposal dates from the time of Pythagoras, and proposes the ratio $81/64$ for the third.
This is far from being a fraction of simple numbers, but it comes about in a natural way from
the Pythagorean system of perfect tuning, as we will see in the next section. In the late 15 th
century the most common tuning system in Europe was called meantone tuning, and in this system
the third was tuned according to the ratio $5/4$. This gives rise to a much more harmonic sound
as the Pythagorean proposal for the third, and that is why we will from now on call it 
the perfect third. The third tuned according to the ratio $81/64$ we will call the Pythagorean
third. Let us consider for a moment the choice of the perfect third. In the case of $c$ as
basic tone this leads to
\be
u(H,e)/u(H,c) = 5/4
\ee
In Table \ref{tableaddingthird} we have added the perfect third.

\begin{table}[h]
\caption{The ratio's of basic, third, fourth, fifth and octave in the case of
perfect and tempered tuning}
\begin{center}
\begin{tabular}{llllll} \hline 
\multicolumn{2}{l}{} & \multicolumn{2}{l}{\bf Harmonic} & \multicolumn{2}{l}{\bf Tempered} \\ \hline
{\it Tone Ratio} & {\it Example} & {\it Freq. Fraction} & {\it Cent} & {\it Freq. Fraction} & {\it Cent} \\ \hline
Basic & $c/c$ & 1 & 0.000 & 1 & 0.000 \\ 
Third & $e/c$ & 5/4 = 1.25 & 386.314 & $(\root 12 \of {2})^4$ = 1.2599 & 400.000\\ 
Fourth & $f/c$ & 4/3 = 1.3333 & 498.044 & $(\root 12 \of {2})^5$ = 1.3348 & 500.000\\ 
Fifth & $g/c$ & 3/2 = 1.5 & 701.955 & $(\root 12 \of {2})^7$ = 1.4983 & 700.000\\ 
Octave & $c_1/c$ & 2  & 1200.000 & 2 & 1200.000 \\ \hline
\end{tabular}
\label{tableaddingthird}
\end{center}
\end{table}
\noindent
We can easily see now the type of problems that perfect tuning is confronted with. Indeed, if
perfect tuning would be possible using the twelve standerd tones of the octave, the tone
$a_b$ should be a perfect third with respect to $e$, and then the tone $c_1$ should be a perfect third with respect to
$a_b$. This would mean however that
\be
u(H,c_1)/u(H,c) = u(H,c_1)/u(H,a_b) \cdot u(H,a_b)/u(H,e) \cdot u(H,e)/u(H,c) = 5/4 \cdot 5/4 \cdot 5/4 = 125/64
\ee
But we know that
\be
u(H,c_1)/u(H,c) = 2 = 128/64
\ee
which means that if we would tune $e$, $a_b$ and $c_1$ following the perfect third ratio, the tone $u(H,c_1)$ would fall
$3/64$ short for a perfect octave. Hence it is impossible to use perfect thirds all the way within the octave if we only have
available the twelve traditional notes. But why has western music been constructed around these twelve traditional notes? Could
this perhaps be the basis of the problem? Let us analyse in the next section the Pythagorean perfect tuning system, because
the reason to put twelve half notes into an octave is to be found in this system.

\section{Twelve Notes in an Octave: The Pythagorean System} \label{Pythagorean}
\noindent The Pythagorean system is a musical
system where the tones are created out of ratios that only contain the numbers 2 and 3. In the 13 th century the
French Academy at Notre Dame decreed that only the Pythagorean tuning was accepted as the correct musical tuning. As a
consequence till the 15 century musical instruments have been tuned following the Pythagorean system.

Let us calculate
the tones that flow naturally out of this Pythagorean approach. We will do this again by starting with
$c$ as the basic note, but obviously the whole process can be repeated with any other choice of basic note.

\subsection{The Basic Pythagorean System}
\noindent
We start from
$u(P,c)$ and have constructed already $u(P,c_1)$ with ratio 2, $u(P,g)$ with ratio $3/2$ and $u(P,f)$ with ratio $2/3$. Since
these ratios only contain numbers 2 and 3, they fit into the Pythagorean system. The perfect third that we constructed
in section \ref{sec:harmonicpossibilities}, corresponding to ratio $5/4$, does not appear in the Pythagorean system.  If we move
an octave plus a fifth further up we have
$u(P,d_3)/u(P,g_1) = 3$, hence $u(P,d_3)/u(P,c) = 3^2 = 9$. Since
$u(P,d_3)/u(P,d) = 2^3 = 8$ we have
\be
u(P,d)/u(P,c) = 9/8
\ee
This gives us the next tone of the Pythagorean system, $u(P,d)$ with ratio 9/8. Sometimes, in analogy with the other
namings, this is also called the Pythagorean second. Let us proceed in a analogous way and calculate $u(P,a)/u(P,c)$. We
have
$u(P,a_4)/u(P,c) = 3^3 = 27$ and
$u(P,a_4)/u(P,a) = 2^4 = 16$ which gives
\be
u(P,a)/u(P,c) = 27/16
\ee
and this ratio is referred to as the Pythagorean sixth.
Further we know that $u(P,e_6)/u(P,c) = 3^4 = 81$ and $u(P,e_6)/u(P,e) = 2^6 = 64$ which gives
\be
u(P,e)/u(P,c) = 81/64
\ee
and we find back the Pythagorean fifth that we mentioned already in the foregoing section. We have $si(7)/c = 3^5 =
243$ and
$u(P,b_7)/u(P,b) = 2^7 = 128$ which gives
\be
u(P,b)/u(P,c) = 243/128 \label{si01}
\ee
and this is called the Pythagorean seventh. We calculated now the complete set of whole tones of the octave and
proceeding with the same method will now start to give us half tones. We have
$u(P,f^\#_9)/u(P,c) = 3^6 = 729$ and
$u(P,f^\#_9)/u(P,f^\#) = 2^9 = 512$ which gives
\be
u(P,f^\#)/u(P,c) = 729/512
\ee
By moving an octave plus a fifth upwards in frequency -- and hence multiplying the basic tone by powers of 3 -- and then
moving down again by octaves -- and hence dividing the basic tone by powers of 2  -- we have gathered 7 of the
notes of the basic octave, namely
$c$,
$d$,
$e$,
$f^\#$,
$g$,
$a$, and
$b$. The note
$f$ we had obtained by moving in the inverse way, first an octave plus a fifth down, which means dividing the basic
tone by powers of 3, and then moving up by octaves, which means multiplying the basic tone by powers of 2. Let us proceed
in this way to get the other tones in the basic octave. We have $u(P,b_{b-4}/u(P,c) = 1/3^2 = 1/9$ and
$u(P,b_b)/u(P,b_{b-4}) = 2^4 = 16$ which gives
\be
u(P,b_b)/u(P,c) = 16/9
\ee
Next we start from $u(P,e_{b-5})/u(P,c) = 1/3^3 = 1/27$ and $u(P,e_b)/u(P,e_{b-5}) = 2^5 = 32$ which gives
\be
u(P,e_b)/u(P,c) = 32/27
\ee
The following step leads us to $a_b$ by noting that $u(P,a_{b-7})/u(P,c) = 1/3^4 = 1/81$ and $u(P,a_b)/u(P,a_{b-7}) = 2^7 =
128$ which gives
\be
u(P,a_b)/u(P,c) = 128/81
\ee
The last note that we have to calculate is $d_b$ and this one we get by noting that $u(P,d_{b-8})/u(P,c) = 1/3^5 = 1/243$
and
$u(P,d_b)/u(P,d_{b-8}) = 2^8 = 256$ which gives
\be
u(P,d_b)/u(P,c) = 256/243
\ee 
Let us present the twelve tones that we have calculated in this way in Table \ref{tablepythagorean}.

\begin{table}[h]
\caption{The twelve notes of the basic octave tuned in the Pythagorean and the tempered system}
\begin{center}
\begin{tabular}{lllllll} \hline 
\multicolumn{3}{l}{} & \multicolumn{2}{l}{\bf Pythagorean} & \multicolumn{2}{l}{\bf Tempered} \\ \hline
&{\it Tone Ratio} & {\it Example} &  {\it Freq. Fraction} & Cent & {\it Freq. Fraction} & {\it Cent} \\ \hline
0 & Basic & $u(c)/u(c)$ & 1 & 0.00 & 1 & 0.00 \\ 
1 & & $u(d_b)/u(c)$ & 256/243 = 1.0534 & 90.22 & $\root 12 \of {2}$ = 1.0594 & 100.00 \\ 
2 & Second & $u(d)/u(c)$ & 9/8 = 1.125 & 203.91 & $(\root 12 \of {2})^2$ = 1.1224 & 200.00\\ 
3 & & $u(e_b)/u(c)$ & 32/27 = 1.1851 & 294.13 &$(\root 12 \of {2})^3$ = 1.1892 & 300.00\\ 
4 & Third & $u(e)/u(c)$ & 81/64 = 1.2656 & 407.82 & $(\root 12 \of {2})^4$ = 1.2599 & 400.00\\ 
5 & Fourth & $u(f)/u(c)$ & 4/3 = 1.3333 & 498.04 & $(\root 12 \of {2})^5$ = 1.3348 & 500.00 \\ 
6 & & $u(f^\#)/u(c)$ & 729/512 = 1.4338 & 611.73 & $(\root 12 \of {2})^6$ = 1.4142 & 600.00 \\ 
7 & Fifth & $u(g)/u(c)$ & 3/2 = 1.5 & 701.95 & $(\root 12 \of {2})^7$ = 1.4983 & 700.00 \\ 
8 & & $u(a_b)/u(c)$ & 128/81 = 1.5802 & 792.17 & $(\root 12 \of {2})^8$ = 1.5874 & 800.00 \\ 
9 & Sixth & $u(a)/u(c)$ & 27/16 = 1.6875 & 905.86 & $(\root 12 \of {2})^9$ = 1.6817 & 900.00\\ 
10 & & $u(b_b)/u(c)$ & 16/9 = 1.7777 & 996.08 & $(\root 12 \of {2})^{10}$ = 1.7817 & 1000.00\\ 
11 & Seventh & $u(b)/u(c)$ & 243/128 = 1.8984 & 1109.77 & $(\root 12 \of {2})^{11}$ = 1.8877 & 1100.00\\ 
12 & Octave & $u(c_1)/u(c)$ & 2 & 1200.00 & 2 & 1200.00 \\ \hline
\end{tabular}
\label{tablepythagorean}
\end{center}
\end{table}
\noindent
We see that the twelve Pythagorean tones are close to the tempered tunes, but they are different. None of them is equal.
The difference of the fifth is 1.95 cent -- the tempered fifth is a little bit too low -- and the difference of the
fourth is 1.96 cent -- the tempered fourth is a little bit too high. These are small differences, and that is the
reason that in the completely equal tempered system, both fifth and fourth still sound harmonically well. Since the
fifht and the fourth are the base of a lot of simple popular music, our completely tempered system does not perform
very badly for this type of music. For the third the difference is 7.82 cent -- the tempered third is too low --
which is much more already. But for the third we have to remark that the perfect third (see Table \ref{tableaddingthird}) is even
lower than the tempered third, with a difference of 13.69 cent. This means that the tempered third comes closer to the
beautiful harmonic sound of the perfect third as it is the case for the Pythagorean third. So also here the tempered
system does not perform very badly, although the difference with the perfect third is still substancial. The tempered
second is too low with 3.91 cent while the tempered $b_b$ is too high with 3.92. This are significant differences. The
tempered seventh is too low with 9.77 cent which is almost a disaster harmonically speaking (or listening).

But, after all, now that we have understood how much trickery is involved to making the harmonic system into a tempered
system, we should be amazed that it works at all. All these twelve tempered tunes are reasonibly close to the
Pythagorean ones. How does this comes about? To shed more light on this we better develop the Pythagorean system to its
completeness. This is what we do in next section.

\subsection{The Extended Pythagorean System}
\noindent
We could have proceeded further along the same lines with the construction of the Pythagorean system. Let us
do this, and for example construct $g_b$. We have $u(P,g_{b-10})/u(P,c) = 1/3^6 = 1/728$ and $u(P,g_b)/u(P,g_{b-10}) =
2^{10} = 1024$ which gives
\be
u(P,g_b)/u(P,c) = 1024/728 = 1.4065
\ee
We see that $u(P,g_b)/u(P,c)$ does not coincide with $u(P,f^\#)/u(P,c)$. Let us
proceed. Next step we find
$b$ again, indeed $b_{-12}$. And we have $u(P,b_{-12})/u(P,c) = 1/3^7 = 1/2187$ and $u(P,b)/u(P,b_{-12}) = 2^{12} = 4096$.
This gives
\be
u(P,b)/u(P,c) = 4096/2187 = 1.8728
\ee
So we have another outcome for $b$ than the
one we had found already in (\ref{si01}). We call this new tone $b^*$ just to distinguish it from $b$. Next comes
$e^*$, and we have
$u(P,e_{-13})/u(P,c) = 1/3^8 = 1/6561$ and $u(P,e)/u(P,e_{-13}) = 2^{13} = 8192$, hence
\be
u(P,e^*)/u(P,c) = 8192/6561 = 1.2485
\ee
Next is $a^*$ and we have $u(P,a_{-15})/u(P,c) = 1/3^9 = 1/19683$ and $u(P,a)/u(P,a_{-15})$ = $2^{15}$ = $32786$, which
gives
\be
u(P,a^*)/u(P,c) = 32786/19683 = 1.6657
\ee
Next comes $d^*$ and we have $u(P,d^*)/u(P,c) = 1/3^{10} = 1/59049$ and $u(P,d)/u(P,d_{-16})$ = $2^{16}$ = $65572$,
which give
\be
u(P,d^*)/u(P,c) = 65572/59049 = 1.1104
\ee
Then comes $g^*$ and we have $u(P,g_{-18})/u(P,c) = 1/3^{11} = 1/177147$ and $u(P,g)/u(P,g_{-18}) = 2^{18} = 262288$,
which gives
\be
u(P,g^*)/u(P,c) = 262288/177147 = 1.4806
\ee 
And finally we arrive at $c^*$. We have $u(P,c_{-19})/u(P,c) = 1/3^{12} = 1/531441$ and $u(P,c)/u(P,c_{-19}) = 2^{19} =
524576$, which gives
\be
u(P,c^*)/u(P,c) = 524576/531441 = 0.9870
\ee
Now we start to see better what the problem is. The $c^*$ that we have constructed has a frequency that is definitely
different from $c$ which we started with. This difference is called the Pythagorean `COMMA' in music theory. The difference
is due to the fact that $2^{19} =  524288$ is a number that is `close' to $3^{12} = 531441$, `but' different. And, it
is indeed impossible that these two numbers would be equal, since one of the numbers is a power of 2, while the other
is a power of 3. 

We can also proceed with the symmetrical construction, let us do this, to see even more clear what
the problem is. The next tone will be $c^\#$. We have $u(P,c^\#_{11})/u(P,c) = 3^7 = 2187$ and $u(P,c^\#_{11})/u(P,c^\#) =
2^{11} = 2048$ which gives
\be
u(P,c^\#)/u(P,c) = 2187/2048 = 1.0678
\ee
Next comes $g^\#$, and we have $u(P,g^\#_{12})/u(P,c) = 3^8 = 6561$ and $u(P,g^\#_{12})/u(P,g^\#) = 2^{12} = 4096$
which gives
\be
u(P,g^\#)/u(P,c) = 6561/4096 = 1.6018
\ee
Next comes $d^\#$, and we have $u(P,d^\#_{14})/u(P,c) = 3^9 = 19683$ and $u(P,d^\#_{14})/u(P,d^\#) = 2^{14} = 16384$
which gives
\be
u(P,d^\#)/u(P,c) = 19683/16384 = 1.2013
\ee
Next comes $a^\#$, and we have $u(P,a^\#_{15})/u(P,c) = 3^{10} = 59049$ and $u(P,a^\#_{15})/u(P,a^\#) = 2^{15} = 32768$
which gives
\be
u(P,a^\#)/u(P,c) = 59049/32768 = 1.8020
\ee
Next comes $f^*$, and we have $u(P,f_{17})/u(P,c) = 3^{11} = 177147$ and $u(P,f_{17})/u(P,f) = 2^{17} = 131072$
which gives
\be
u(P,f^*)/u(P,c) = 177147/131072 = 1.3515
\ee
And then we are again back at $c$, notated this time $c^{**}$. We have $u(P,c_{19})/u(P,c) = 3^{12} = 531441$ and
$u(P,c_{19})/u(P,c) = 2^{19} = 524288$ which gives
\be
u(P,c^{**})/u(P,c) = 531441/524288 = 1.0136
\ee
This means that we have gathered in this way 25 different notes that lay within the octave interval between $c$
and $c_1$. We represent this situation in Table \ref{tablepythagoreanextended}.

\begin{table}[h]
\caption{The 25 tones that a further calculation of the Pythagorean system gives rise to}
\begin{center}
\begin{tabular}{lllll|} \hline 
& {\it Tone Ratio} & {\it Example} & {\it Pythagorean Fraction} & {\it Cent} \\ \hline
1 & & $u(c^*)/u(c)$ & $2^{19} \cdot 3^{-12} = 524576/531441$ = 0.9870 & -23.46\\
2 & Basic & $u(c)/u(c)$ & 1 & 0.00 \\ 
3 & & $u(c^{**})/u(c)$ & $2^{-19} \cdot 3^{12} = 531441/524288$ = 1.0136 & 23.46 \\ 
4 & & $u(d_b)/u(c)$ & $2^8 \cdot 3^{-5} = 256/243$ = 1.0534 & 90.22\\ \hline
5 & & $u(c^\#)/u(c)$ & $2^{-11} \cdot 3^7 = 2187/2048$ = 1.0678 & 113.68 \\ 
6 & & $u(d^*)/u(c)$ & $2^{18} \cdot 3^{-11} = 65572/59049$ = 1.1104 & 180.44 \\ \hline
7 & Second & $u(d)/u(c)$ & $2^{-3} \cdot 3^2 = 9/8$ = 1.125 & 203.91\\ 
8 & & $u(e_b)/u(c)$ & $2^5 \cdot 3^{-3} = 32/27$ = 1.1851 & 294.13 \\ \hline
9 & & $u(d^\#)/u(c)$ & $2^{-14} \cdot 3^9 = 19683/16384$ = 1.2013 & 317.59 \\ 
10 & & $u(e^*)/u(c)$ & $2^{13} \cdot 3^{-8} = 8192/6561$ =  1.2485 & 384.35 \\ \hline
11 & Pyth. Third & $u(e)/u(c)$ & $2^{-6} \cdot 3^4 = 81/64$ = 1.2656 & 407.82 \\ 
12 & Fourth & $u(f)/u(c)$ & $2^2 \cdot 3^{-1} = 4/3$ = 1.3333 & 498.04 \\ \hline
13 & & $u(f^*)/u(c)$ & $2^{-17} \cdot 3^{11} = 177147/131072$ = 1.3515 & 521.50 \\ 
14 & & $u(g_b)/u(c)$ & $2^{10} \cdot 3^{-6} = 1024/728$ = 1.4065 & 588.26 \\ \hline
15 & & $u(f^\#)/u(c)$ & $2^{-9} \cdot 3^6 = 729/512$ = 1.4338 & 611.73 \\ 
16 & & $u(g^*)/u(c)$ &  $2^{18} \cdot 3^{-11} = 262288/177147$ = 1.4806 & 678.49 \\ \hline
17 & Fifth & $u(g)/u(c)$ & $2^{-1} \cdot 3^1 = 3/2$ = 1.5 & 701.95 \\ 
18 & & $u(a_b)/u(c)$ & $2^7 \cdot 3^{-4} = 128/81$ = 1.5802 & 792.17 \\ \hline
19 & & $u(g^\#)/u(c)$ & $2^{-12} \cdot 3^8 = 6561/4096$ = 1.6018 & 815.64 \\ 
20 & & $u(a^*)/u(c)$ & $2^{15} \cdot 3^{-9} = 32786/19683$ = 1.6657 & 882.40 \\ \hline
21 & Sixth & $u(a)/u(c)$ & $2^{-4} \cdot 3^3 = 27/16$ = 1.6875 & 905.86 \\ 
22 & & $u(b_b)/u(c)$ & $2^4 \cdot 3^{-2} = 16/9$ = 1.7777 & 996.08 \\ \hline
23 & & $u(a^\#)/u(c)$ & $2^{-15} \cdot 3^{10} = 59049/32768$ = 1.8020 & 1019.55 \\ 
24 & & $u(b^*)/u(c)$ & $2^{12} \cdot 3^{-7} = 4096/2187$ = 1.8728 & 1086.31 \\ \hline
25 & Seventh & $u(b)/u(c)$ & $2^{-7} \cdot 3^5 = 243/128$ = 1.8984 & 1109.77 \\ 
26 & Octave & $u(c_1)/u(c)$ & 2 & 1200.00\\ \hline
\end{tabular}
\end{center}
\label{tablepythagoreanextended}
\end{table}
\noindent
How to choose now the tones that we will finally use to tune
our musical instrument? And here the problem arrives. A good choice would be to first choose the tones that have the
most simple ratio towards the basic tone. In case of basic note $c$, we would then choose in the following way.
There is one note, namely $c_1$, that is our first choice. Then there are three notes $g$, $f$ and $d$
that only have integers smaller than 10 in their fraction, so they come next. There are 4 tones, $e_b$, $e$
$a$ and
$b_b$ that have integers smaller than 100 in their fractions, so they come next. There are 4 tones, $d_b$,
$f^\#$, $a_b$ and $b$, that have integers smaller than 1000 in their fractions, so next we choose them.
And this gives us already all the twelve common notes that also exist in a tempered tuning system. So we could choose just
to drop the others, and tune our instrument with these 12 most simple fractions. These are the ones that we have
presented in Table \ref{tablepythagorean}. That is also how an instrument is tuned in the Pythagorean system. The problem is however that this
choice `depends' on the basic tone that we start with. So we could make this choice for the $c$ octave, but the same
way of choosing would lead to a different set of tones for another octave. And that is the reason that by this method is
it not possible to realize a Pythagorean tuning system that is valid independent of the basic note that is chosen. If we
tune following our Pythagorean calculus with basic note $c$ our music instrument will be well tuned for music played
in this basic tone. But if on this instrument we want to play in another basic tone (another key), we would have to tune
all over again. This means that the Pythagorean tuning, as we have introduced it here, and as it has been introduced in
all of history, is not resistent against transportation between different keys. How can we proceed?

\subsection{Pythagorean Tuning and Different Keys}
\noindent If the problem with the Pythagorean tuning would reveal itself if transportation of a music piece is made
to another key, it would perhaps be still possible to cope with it. Not in the old times, were it was technically not
possible to each time tune an instrument again when a music piece would be played in a different key. But with modern
musical instruments like synthesizers, it is possible to tune the instrument electronically again just by pushing a
button. Within the community of musicians that are interested in microtonal music, which is music using
types of tuning systems different from the completely tempered system that a standard synthesizer is tuned in, these
possibilities have been explored. So this would make it technically possible to adopt a new tuning each time that a music piece is played in a different key. The real hard problem of perfect tuning is however not this one. In more
complex musical works keys can change within the works itself. Now even here one might think that it would be possible
to change the tuning at the moment that a change of key appears, and so a solution would exist when these works are
played by such a sophisticated synthesizer. The problem is however that this only would be a solution for musical works
where abruptly a change of key appears, so when really a new theme in a new key starts at a specific moment. Real
complex works that explore harmony in depth will however contain gradual change, more a kind of slow deviation out of
the old key towards a new key. And then it would become a matter of taste or purely subjective decision of the musician
when to change tuning. There does not seem to be a systematic way to do this as long as we stick with the type of
Pythagorean tuning as the one that we have introduced here, and hence the one that is historically known. It is this
problem that gets solved in depth by the introduction of the intrinsic harmonic spaces that we present in this article.
To be able to prove this, we first have to introduce these intrinsic harmonic spaces, and that is wat we will do in the
next section. 

\section{Intrinsic Harmonic Spaces}
\noindent
In this section we introduce an intrinsic harmonic space for the Pythagorean system. As we will show it is a two
dimensional vector space. In section \ref{section04} we show how the procedure is general and can be elaborated for any
kind of harmonic tuning system, also the more elaborated ones than the Pythagorean system. In general we will then
need a vector space of higher dimensions, the number of dimensions being equal to the number of prime numbers
different from 1 that are allowed for the ratio's of the frequencies of different tones.

\subsection{The Pythagorean Intrinsic Harmonic Space}
\noindent
For the Pythagorean intrinsic harmonic space we consider ratio's between frequencies of tones that only contain prime
numbers 1, 2 and 3. Let us look again at equation (\ref{v/u}), which we have introduced for arbitrary tones $u$ and $v$.
In principle, for arbitrary tones, $k$ can be any real positive number. If we want to analyse Pythagorean harmonized
tones,
$k$ must be the product of a power of 2 and a power of 3. This means that we can always write, for $m, n \in \integ$
\be
k = 2^m \cdot 3^n \label{mn}
\ee
These numbers $m$, and $n$ represent in a unique way the relation between $u$ and $v$, because there is only
one way to write an arbitrary number that is a product of powers of 2 and powers of 3. We can thus take
$(m, n)$ as a representative of this relation. 

If we want to remain general, we want to include all the tones
that can be formed in this way, starting from a basic tone. We shall see that there is an intrinsic way to
represent these tones in a two dimensional vector space. Indeed, suppose that we take the basic tone $u$, and
represent this basic tone by the $(0, 0)$ vector of the two dimensional vector space (the origin). Then we can
represent any other tone that is related to $v$ by equation (\ref{v/u}) by means of the point $(m, n)$ where $m$
and $n$ are the two whole numbers that appear in equation (\ref{mn}). Let us give some examples. If $u$ is $u(P,c)$,
then in this two dimensional vector space
$u(P,c)$ will represented by the centre $(0, 0)$. The tone $u(P,c_1)$ will be represented by the point $(1, 0)$. Indeed:
\begin{equation}
u(P,c_1)/u(P,c) = 2 = 2^1 \cdot 3^0
\end{equation}
The tone $u(P,g_1)$ is represented by the point $(0, 1)$ because
\begin{equation}
u(P,g_1)/u(P,c) = 3 = 2^0 \cdot 3^1
\end{equation}
Hence $u(P,c_1)$ and $u(P,g_1)$ form a canonical base of our vector space. The tone $u(P,c_{-1})$ is represented by the point
$(-1, 0)$ because
\begin{equation}
u(P,c_{-1})/u(P,c) = 1/2 = 2^{-1} \cdot 3^0
\end{equation}
and the tone $u(P,f_{-2})$ is represented by the point $(0, -1)$. Indeed it is the tone $u(P,c)$ which is a fifth
higher than the tone $u(P,f_{-2})$, hence
\begin{equation}
u(P,f_{-2})/u(P,c) = 1/3 = 2^0 \cdot 3^{-1}
\end{equation}
What is the place of the tone $u(P,g)$ for example in our vector space? We have
\begin{equation}
u(P,g)/u(P,c) = 3/2 =  2^{-1} \cdot 3^1
\end{equation}
and therefore $u(P,g)$ will be represented by the point $(-1, 1)$. What will be the point representing $u(P,f)$?

\vskip 0.5 cm
\begin{center}
\begin{picture}(175,220)
\linethickness{0.1pt}
\put(80,110){\circle*{5}}
\multiput(0,20)(40,0){5}{\line(0,1){180}}
\multiput(-10,30)(0,40){5}{\line(1,0){180}}
\put(63,101){\makebox[1 cm][l]{\smallroman (0,0)}}
\put(103,101){\makebox[1 cm][l]{\smallroman (1,0)}}
\put(120,110){\circle*{5}}
\put(80,150){\circle*{5}}
\put(40,150){\circle*{5}}
\put(80,70){\circle*{5}}
\put(120,70){\circle*{5}}
\put(160,70){\circle*{5}}
\put(100,61){\makebox[1 cm][l]{\smallroman (1,-1)}}
\put(60,61){\makebox[1 cm][l]{\smallroman (0,-1)}}
\put(63,141){\makebox[1 cm][l]{\smallroman (0,1)}}
\put(20,141){\makebox[1 cm][l]{\smallroman (-1,1)}}
\put(140,61){\makebox[1 cm][l]{\smallroman (2,-1)}}
\thicklines
\put(-30,110){\vector(1,0){220}}
\put(80,0){\vector(0,1){220}}
\put(180,117){\makebox[1 cm][l]{$\scriptstyle m$}}
\put(87,213){\makebox[1 cm][l]{$\scriptstyle n$}}
\put(83,154){\makebox[1 cm][l]{$\scriptstyle g_1$}}
\put(43,154){\makebox[1 cm][l]{$\scriptstyle g$}}
\put(83,114){\makebox[1 cm][l]{$\scriptstyle c$}}
\put(83,74){\makebox[1 cm][l]{$\scriptstyle f_{-2}$}}
\put(123,114){\makebox[1 cm][l]{$\scriptstyle c_1$}}
\put(123,74){\makebox[1 cm][l]{$\scriptstyle f_{-1}$}}
\put(163,74){\makebox[1 cm][l]{$\scriptstyle f$}}
\end{picture}
\end{center}
\begin{center}
\begin{minipage}{8 cm}
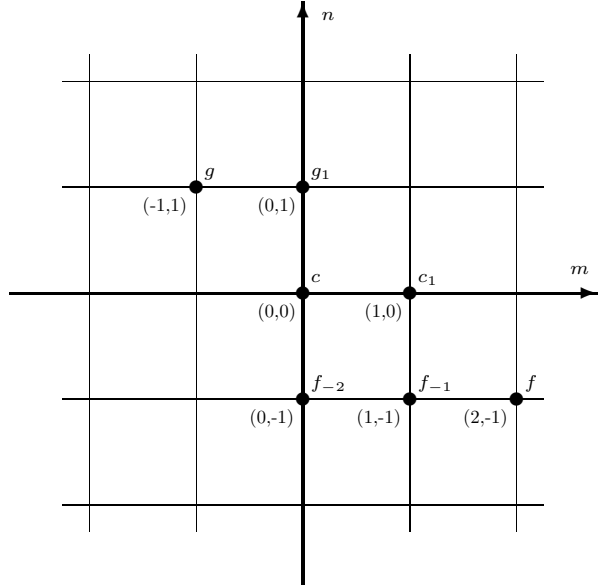
\mediumroman \baselineskip 9 pt Figure 1: A representation of the tones $\scriptstyle c$,
$\scriptstyle c_1$, $\scriptstyle g$, $\scriptstyle g_1$, $\scriptstyle f_{-1}$,
$\scriptstyle f_{-2}$ in the two dimensional vectorspace
\end{minipage}
\end{center}
\vskip 0.5 cm
We have
\begin{equation}
u(P,f_{-2})/u(P,c) = 1/3 =  2^0 \cdot 3^{-1}
\end{equation}
which shows that the point $(0,-1)$ represents $u(P,f_{-2})$. From this follows that
\begin{equation}
u(P,f_{-1})/u(P,c) = 2/3 =  2^1 \cdot 3^{-1}
\end{equation}
which shows that the point $(1, -1)$ represents $u(P,f_{-1})$, and as a consequence
\begin{equation}
u(P,f)/u(P,c) = 4/3 =  2^2 \cdot 3^{-1}
\end{equation}
which shows that $u(P,f)$ is represented by the point $(2, -1)$. In Figure 1 we
represent this situation.

Let us consider an arbitrary tone $v$, represented by the vector $(m, n)$. This means that $v/u = 2^m
\cdot 3^n$. Consider now another tone $w$, represented by the vector $(k, l)$, which means that $w/u =
 2^k \cdot 3^l$. Let us calculate the relation between $v$ and $w$. We have $w/v = w/u \cdot u/v
= 2^k \cdot 3^l \cdot 2^{-m} \cdot 3^{-n} = 2^{k-m} \cdot 3^{l-n}$. This means that $w$ is
represented towards $v$ by the vector $(k-m, l-n) = (k, l) - (m, n)$. This proves that our representation is
intrinsic. When we choose another point as origin of our two dimensional vector space, for example the point $(m,
n)$, then the new vector $(m, n) - (k, l)$ represents $w$ with respect to $v$.

\subsection{The Octave Projection}

\noindent
We want to investigate now how we get back our traditional Pythagorean system, the one we have presented in section
\ref{Pythagorean}. This means that we have to investigate which tones are in which octave. This we can see in the
following way. A tone is in the octave
$[c, c_1]$ if its relation to
$c$ is in de interval
$[1,2]$. So we can say in general that a tone is in the octave $[(m, 0), (m+1, 0)]$ if its relation to the basic tone
$(0, 0)$ is in the interval $[2^m, 2^{m+1}]$. Let us consider then a general tone represented by $(x, y)$.
The relation between this tone and the basic tone is given by  $2^x \cdot 3^y$. So to have this tone in
the octave $[(m, 0), (m+1, 0)]$ we must have:
\begin{equation}
2^m \le 2^x \cdot 3^y \le 2^{m+1}
\end{equation}
This is equivalent to
\begin{equation}
m\log2 \le x\log2 + y\log3 \le (m+1)\log2
\end{equation}
Let us draw the lines given by the equations
\begin{equation}
x\log2 + y\log3 = m\log2 \label{octaveline} 	
\end{equation}
Suppose that $y = 0$ then $x = m$, which means that the straight lines that are determined by equation
(\ref{octaveline}) cut the $x$-axis in the point $m$. On the other hand, when $x = 0$, then
$y = m \cdot \log2/\log3$. This means that these straight lines cut the $y$-axis in the points $m \cdot
\log2/\log3$. The direction coefficient of all the lines is given by $-\log2/\log3 = -0.6309$. In Figure 2 we have drawn the
tones that are contained in the octave $[c, c_1]$, and are between the two lines, one with $m = 0$ and the other with
$m=1$.

In Figure 2 we represent tones as they have been projected along a line with direction coefficient $\log2/\log3 = -0.6309$
onto the
$x$-axis.
To be able to find back the cent value for each projected tone on the interval $[(0,0), (0,1)]$ we have to multiply the
unit of the
$x$-axis with $1200$ such that $c_1$ is now in the point $(0, 1200)$. Let us calculate the values of the projections on
the $x$-axis of the different tones of octave $[c, c_1]$. For $g$ we get
\be
proj(g) = (\log3 - \log2) \cdot 1200/\log2  = 701.955
\ee
For $d$ we have
\be
proj(d) = (2 \log3 - 3\log2) \cdot 1200/\log2 = 203.910
\ee
For $a$ we have
\be
proj(a) = (3\log3 - 4\log2) \cdot 1200/\log2 = 905.865
\ee
For $e$ we have
\be
proj(e) = (4\log3 - 6\log2) \cdot 1200/\log2 = 407.820
\ee
For $b$ we have
\be
proj(b) = (5\log3 - 7\log2) \cdot 1200/\log2 = 1109.775
\ee
For $f^\#$ we have
\be
proj(f^\#) = (6\log3 - 9\log2) \cdot 1200/\log2 = 611.730
\ee

\vskip 0.5 cm
\begin{center}
\begin{picture}(460,320)
\linethickness{0.1pt}
\put(225,159){\circle*{5}}
\multiput(5,2)(22,0){21}{\line(0,1){314}}
\multiput(2,5)(0,22){15}{\line(1,0){446}}
\put(208,150){\makebox[1 cm][l]{\smallroman (0,0)}}
\put(227,152){\makebox[1 cm][l]{$\scriptstyle c$}}
\put(250,164){\makebox[1 cm][l]{\smallroman (1,0)}}
\put(240,164){\makebox[1 cm][l]{$\scriptstyle c_1$}}
\put(247,159){\circle*{5}}
\put(203,181){\circle*{5}}
\put(207,175){\makebox[1 cm][l]{$\scriptstyle g$}}
\put(159,203){\circle*{5}}
\put(163,197){\makebox[1 cm][l]{$\scriptstyle d$}}
\put(137,225){\circle*{5}}
\put(141,219){\makebox[1 cm][l]{$\scriptstyle a$}}
\put(93,247){\circle*{5}}
\put(97,241){\makebox[1 cm][l]{$\scriptstyle e$}}
\put(71,269){\circle*{5}}
\put(75,263){\makebox[1 cm][l]{$\scriptstyle b$}}
\put(27,291){\circle*{5}}
\put(31,285){\makebox[1 cm][l]{$\scriptstyle f^\#$}}
\put(269,137){\circle*{5}}
\put(273,131){\makebox[1 cm][l]{$\scriptstyle f$}}
\put(313,115){\circle*{5}}
\put(317,109){\makebox[1 cm][l]{$\scriptstyle b_b$}}
\put(335,93){\circle*{5}}
\put(339,87){\makebox[1 cm][l]{$\scriptstyle e_b$}}
\put(379,71){\circle*{5}}
\put(383,65){\makebox[1 cm][l]{$\scriptstyle a_b$}}
\put(401,49){\circle*{5}}
\put(405,43){\makebox[1 cm][l]{$\scriptstyle g_b$}}
\qbezier(203,181)(237.86,159)(237.86,159)
\qbezier(159,203)(228.74,159)(228.74,159)
\qbezier(137,225)(241.60,159)(241.60,159)
\qbezier(93,247)(232.25,159)(232.25,159)
\qbezier(71,269)(245.35,159)(245.35,159)
\qbezier(27,291)(236.21,159)(236.21,159)
\qbezier(269,137)(234.13,159)(234.13,159)
\qbezier(313,115)(243.26,159)(243.26,159)
\qbezier(335,93)(230.39,159)(230.39,159)
\qbezier(379,71)(239.52,159)(239.52,159)
\qbezier(401,49)(226.65,159)(226.65,159)
\qbezier(445,27)(235.78,159)(235.78,159)
\thicklines

\end{picture}
\end{center}
\begin{center}
\begin{minipage}{14 cm}
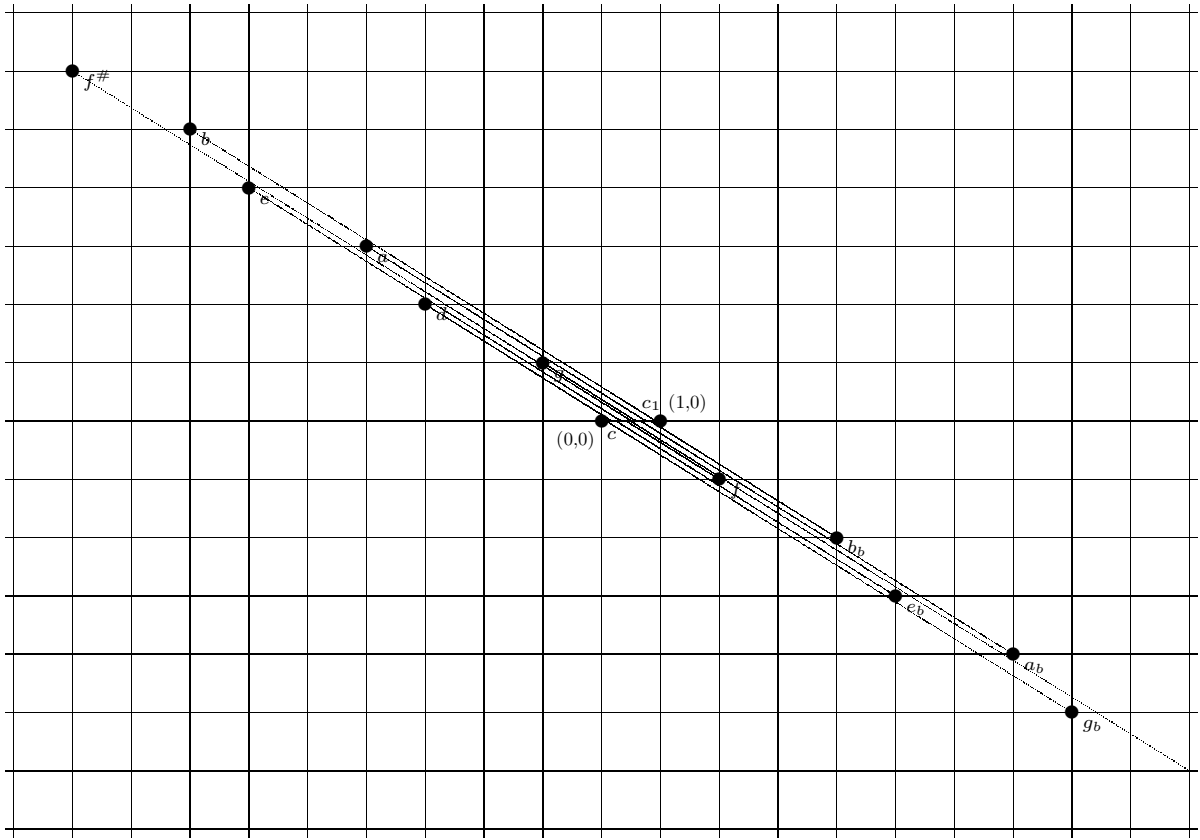
\mediumroman \baselineskip 9 pt Figure 2: The Pythagorean tones in the vector space representation and the octave
projection of these tones
\end{minipage}
\end{center}
\vskip 0.2 cm
For $f$ we have
\be
proj(f) = (2\log2 - \log3) \cdot 1200/\log2 = 498.045
\ee
For $b_b$ we have
\be
proj(b_b) = (4\log2 - 2\log3) \cdot 1200/\log2 = 996.090
\ee
For $e_b$ we have
\be
proj(e_b) = (5\log2 - 3\log3) \cdot 1200/\log2 = 294.135
\ee
For $a_b$ we have
\be
proj(a_b) = (7\log2 - 4\log3) \cdot 1200/\log2 = 792.180
\ee
For $d_b$ we have
\be
proj(d_b) = (8\log2 - 5\log3) \cdot 1200/\log2 = 90.225
\ee
And finally for $g_b$ we have
\be
proj(g_b) = (10\log2 - 6\log3) \cdot 1200/\log2 = 588.270
\ee
We see that the projections slowly deviate more and more from the temperated tones, that correspond in the
representation of Figure 2 with sums of pieces of $1/12$ of the interval $[c, c_1]$. For example the
deviation of $g$ is only $1.956$ cent, while the deviation of $f^\#$ is already $11.730$ cent. In Table \ref{tableoctaveprojection}
we represent the different tones and the measures of their intervals.

\begin{table}[h]
\caption{A representation of the octave
projection giving rise to the tones within the octave interval as they appear in the traditional Pythagorean system}
\begin{center}
\begin{tabular}{llllll} \hline
{\it Vector} & {\it Tone Ratio} & {\it Example} & {\it Projection Value} & {\it Deviation} \\ \hline
 (0, 0) & Basic & $c$ & 0.000 & 0.000 \\ 
 (8, -5) & & $d_b$ & 90.225 & 9.775 \\ 
 (-3, 2) & Second & $d$ & 203.910 & 3.910 \\ 
 (5, -3) & Minor Third & $e_b$ & 294.135 & 5.865\\
 (-6, 4) & Third & $e$ & 407.820 & 7.820 \\ 
 (2, -1) & Fourth & $f$ & 498.045 & 1.955 \\ 
 (10, -6) & & $g_b$ & 588.270 & 11.730 \\ 
 (-9, 6) & & $f^\#$ & 611.730 & 11.730 \\ 
 (-1, 1) & Fifth & g & 701.955 & 1.955 \\ 
 (7, -4) & & $a_b$ & 792.180  & 7.820 \\ 
 (4, -2) & Sixth  & $a$ & 905.865 & 5.865 \\ 
 (4, -2) & & $b_b$ & 996.090 & 3.910 \\ 
 (-7, 5) & Seventh & $b$ & 1109.775 & 9.775 \\ 
 (1, 0) & Octave & $c_1$ & 1200.000 & 0.000 \\ \hline
\end{tabular}
\label{tableoctaveprojection}
\end{center}
\end{table}

\subsection{The Intrinsic Musical Structure}
\noindent Let us show by means of a concrete
example the difference between the intrinsic musical structure that we can make correspond to each musical piece
in our intrinsic harmonic space and the traditional way that music pieces are represented.

We choose a part of the
leading theme of Beethoven's Fifth Symphony of which the consecutive notes are the following: $e$, $f$, $g$,
$g$,
$f$,
$e$, $d$, $c$, $c$, $d$, $e$, $e$, $d$, $d$.

Let us write this in our vector
representation: (-6, 4), (2, -1), (-1, 1), (-1, 1), (2, -1), (-6, 4), (-3, 2), (0, 0), (0, 0), (-3, 2), (-6, 4), (-6,
4), (-3, 2), (-3, 2). We also want to calculate the vectors that carry us in the vector space from one tone to
another, because it is the collection of these vectors that determines completely the harmonic structure of the
musical pattern formed by this leading theme of Beethoven's Fifth Symphony. Consider for example the first tone and
second tone of the theme, which are $e$ and $f$ represented by the vectors $(-6, 4)$ and $(2, -1)$. It is the
vector $(8, -5) = (2, -1) - (-6, 4)$ that carries the $e$ to the
$f$. In a similar way, it is the vector $(-3, 2) = (-1, 1) - (2, -1)$ that carries the second tone $f$ to the
third tone $g$ of the theme. In this way we can represent the tone by these `difference' vectors in a way that
is independent of the starting tone. In Table \ref{tablebeethoventheme} we have represented tones, vectors and difference vectors for
the Beethoven theme.

\begin{table}[h]
\caption{The Beethoven theme with the first column the vectors, the second column the names of the tones, the  third column the difference vectors and the fourth column the actions as named in the traditional way.}
\begin{center}
\begin{tabular}{lllll} \hline
Vector & Tone & Difference Vector & Action \\ \hline
 (-6, 4) & $e$ & (8, -5) = (2, -1) - (-6, 4)  & half tone up \\ 
 (2, -1) & $f$ & (-3, 2) = (-1, 1) - (2, -1)& tone up \\ 
 (-1, 1) & $g$ & (0, 0) = (-1, 1) - (-1, 1) & stay \\ 
 (-1, 1) &  $g$ & (3, -2) = (2, -1) - (-1, 1) & tone down \\ 
 (2, -1) & $f$ & (-8, 5) =  (-6, 4) - (2, -1)& half tone down \\ 
 (-6, 4) & $e$ & (3, -2) = (-3, 2) - (-6, 4) & tone down \\ 
 (-3, 2) & $d$ & (3, -2) = (0, 0) - (-3, 2)& tone down \\ 
 (0, 0) & $c$ & (0, 0) = (0, 0) - (0, 0) & stay \\ 
 (0, 0) & $c$ & (-3, 2) = (-3, 2) - (0, 0) & tone up \\ 
(-3, 2) & $d$ & (-3, 2) = (-6, 4) - (-3, 2)  & tone up \\ 
 (-6, 4) & $e$ & (0, 0) = (-6, 4) - (-6, 4)& stay \\ 
 (-6, 4) & $e$ & (3, -2) = (-3, 2) - (-6, 4)& tone down \\ 
 (-3, 2) & $d$ & (0, 0) = (-3, 2) - (-3, 2)& stay \\ 
 (-3, 2) &  $d$ &  &  \\ \hline
\end{tabular}
\label{tablebeethoventheme}
\end{center}
\end{table}
In Figure 3 we have drawn the intrinsic musical pattern corresponding to this theme.
\begin{center}
\setlength{\unitlength}{0.7 pt}
\begin{picture}(460,320)
\linethickness{0.1pt}
\multiput(5,2)(22,0){21}{\line(0,1){314}}
\multiput(2,5)(0,22){15}{\line(1,0){446}}
\put(225,159){\circle*{5}} 
\put(225,150.6){\circle{16.8}} 
\put(247,159){\circle*{5}} 
\put(203,181){\circle*{5}} 
\put(203,189.4){\circle{16.8}} 
\put(159,203){\circle*{5}} 
\put(137,225){\circle*{5}} 
\put(93,247){\circle*{5}} 
\put(87,253){\circle{16.8}} 
\put(71,269){\circle*{5}} 
\put(27,291){\circle*{5}} 
\put(269,137){\circle*{5}} 
\put(313,115){\circle*{5}} 
\put(335,93){\circle*{5}} 
\put(379,71){\circle*{5}} 
\put(401,49){\circle*{5}} 
\put(445,27){\circle*{5}} 
\qbezier(93,247)(269,137)(269,137)
\qbezier(269,137)(203,181)(203,181)
\qbezier(93,247)(159,203)(159,203)
\qbezier(225,159)(159,203)(159,203)
\thicklines
\put(0,159){\vector(1,0){454}}
\put(225,0){\vector(0,1){322}}
\put(451,164){\makebox[1 cm][l]{$\scriptstyle m$}}
\put(229,322){\makebox[1 cm][l]{$\scriptstyle n$}}
\put(209,195){\makebox[1 cm][l]{$\scriptstyle sol$}}
\put(144,193){\makebox[1 cm][l]{$\scriptstyle re$}}
\put(140,232){\makebox[1 cm][l]{$\scriptstyle la$}}
\put(75,235){\makebox[1 cm][l]{$\scriptstyle mi$}}
\put(75,275){\makebox[1 cm][l]{$\scriptstyle si$}}
\put(6,280){\makebox[1 cm][l]{$\scriptstyle fa^\#$}}
\put(204,143){\makebox[1 cm][l]{$\scriptstyle do$}}
\put(252,128){\makebox[1 cm][l]{$\scriptstyle fa$}}
\put(249,165){\makebox[1 cm][l]{$\scriptstyle do$}}
\put(316,120){\makebox[1 cm][l]{$\scriptstyle si_b$}}
\put(315,83){\makebox[1 cm][l]{$\scriptstyle mi_b$}}
\put(382,76){\makebox[1 cm][l]{$\scriptstyle la_b$}}
\put(384,40){\makebox[1 cm][l]{$\scriptstyle re_b$}}
\put(424,18){\makebox[1 cm][l]{$\scriptstyle sol_b$}}
\end{picture}
\end{center}
\begin{center}
\begin{minipage}{9.5 cm}
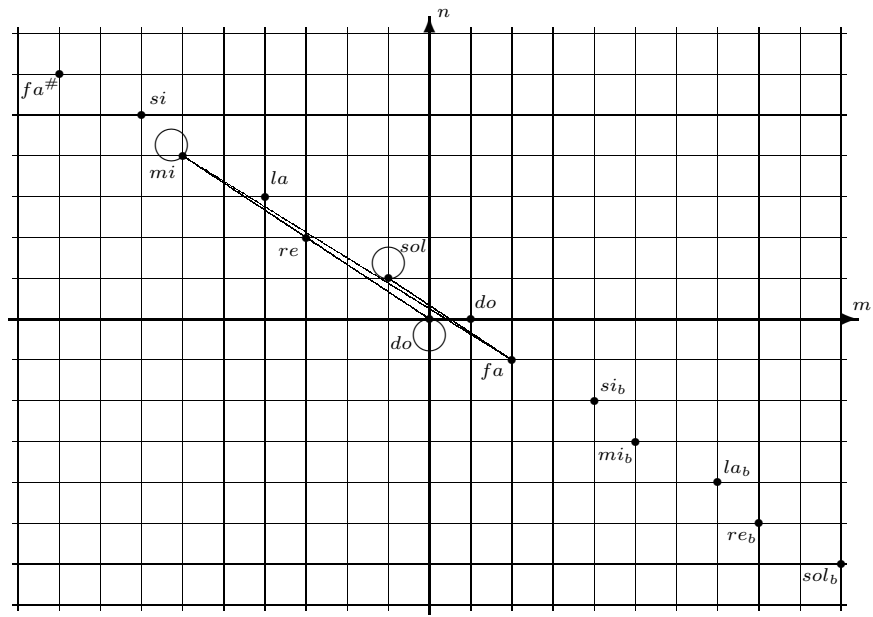
\mediumroman \baselineskip 9 pt Figure 3: The intrinsic musical pattern corresponding to the Beethoven theme
\end{minipage}
\end{center}
We can transpose this pattern all over the two dimensional vector space, and each time it will represent the
Beethoven theme played in a different key. This is exactly also what the tempered tuning system allows. But our scheme
allows this transposition within a full harmonic system. In our scheme the transposition takes place while at the same
time the full harmony of the Pythagorean system is conserved. Let us transpose the theme for example such that the
starting note is $b_b$, hence vector $(4, -2)$.
\vspace{-1.4 cm}
\begin{center}
\setlength{\unitlength}{0.7 pt}
\hspace{2.5 cm} \begin{picture}(460,320)
\linethickness{0.1pt}
\put(225,159){\circle*{5}} 
\put(225,150.6){\circle{16.8}} 
\put(203,181){\circle*{5}} 
\put(203,189.4){\circle{16.8}} 
\put(159,203){\circle*{5}} 
\put(93,247){\circle*{5}} 
\put(87,253){\circle{16.8}} 
\put(269,137){\circle*{5}} 
\qbezier(93,247)(269,137)(269,137)
\qbezier(269,137)(203,181)(203,181)
\qbezier(93,247)(159,203)(159,203)
\qbezier(225,159)(159,203)(159,203)
\thicklines
\put(209,195){\makebox[1 cm][l]{$\scriptstyle sol$}}
\put(144,193){\makebox[1 cm][l]{$\scriptstyle re$}}
\put(75,235){\makebox[1 cm][l]{$\scriptstyle mi$}}
\put(204,143){\makebox[1 cm][l]{$\scriptstyle do$}}
\put(252,128){\makebox[1 cm][l]{$\scriptstyle fa$}}
\end{picture}
\end{center}
\begin{center}
\vspace{-2.7 cm}\begin{minipage}{5 cm}
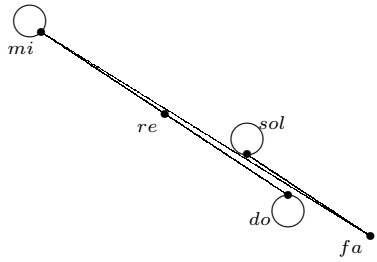
\mediumroman \baselineskip 9 pt Figure 4: The Beethoven pattern on itself
\end{minipage}
\end{center}
In Table \ref{tabletranspositionbeethoven} we present this situation and in Figure 4 this new
situation is represented in the vectorspace.

\begin{table}[h]
\caption{A transposition of the Beethoven theme}
\begin{center}
\begin{tabular}{lllll} \hline
Vector & Tone & Difference Vector & Action \\ \hline
 (4, -2) & $b_b$ & (8, -5)  & half tone up \\ 
 (12, -7) & $b$ & (-3, 2) & tone up \\ 
 (9, -5) & $d_b$ & (0, 0) & stay \\ 
 (9, -5) &  $d_b$ & (3, -2) & tone down \\ 
 (12, -7) & $b$ & (-8, 5) & half tone down \\ 
 (4, -2) & $b_b$ & (3, -2) & tone down \\ 
 (7, -4) & $a_b$ & (3, -2) & tone down \\ 
 (10, -6) & $g_b$ & (0, 0) & stay \\ 
 (10, -6) & $g_b$ & (-3, 2) & tone up \\ 
(7, -4) & $a_b$ & (-3, 2)  & tone up \\ 
 (4,-2) & $b_b$ & (0, 0) & stay \\ 
 (4, -2) & $b_b$ & (3, -2) & tone down \\ 
 (7, -4) & $d$ & (0, 0) & stay \\ 
 (7, -4) &  $d$ &  &  \\ \hline
\end{tabular}
\label{tabletranspositionbeethoven}
\end{center}
\end{table}

\begin{center}
\setlength{\unitlength}{0.7 pt}
\begin{picture}(460,320)
\linethickness{0.1pt}
\multiput(5,2)(22,0){21}{\line(0,1){314}}
\multiput(2,5)(0,22){15}{\line(1,0){446}}
\put(225,159){\circle*{5}} 
\put(445,18.6){\circle{16.8}} 
\put(247,159){\circle*{5}} 
\put(203,181){\circle*{5}} 
\put(423,57.4){\circle{16.8}} 
\put(159,203){\circle*{5}} 
\put(137,225){\circle*{5}} 
\put(93,247){\circle*{5}} 
\put(307,121){\circle{16.8}} 
\put(71,269){\circle*{5}} 
\put(27,291){\circle*{5}} 
\put(269,137){\circle*{5}} 
\put(313,115){\circle*{5}} 
\put(335,93){\circle*{5}} 
\put(379,71){\circle*{5}} 
\put(401,49){\circle*{5}} 
\put(445,27){\circle*{5}} 
\put(489,5){\circle*{5}} 
\put(423,49){\circle*{5}} 
\put(203,189.4){\circle{16.8}} 
\put(225,150.6){\circle{16.8}} 
\put(87,253){\circle{16.8}} 
\qbezier(313,115)(489,5)(489,5) 
\qbezier(269,137)(203,181)(203,181)
\qbezier(313,115)(445,27)(445,27)
\qbezier(93,247)(159,203)(159,203)
\qbezier(225,159)(159,203)(159,203)
\qbezier(489,5)(423,49)(423,49)
\qbezier(269,137)(93,247)(93,247)
\thicklines
\put(0,159){\vector(1,0){454}}
\put(225,0){\vector(0,1){322}}
\put(451,164){\makebox[1 cm][l]{$\scriptstyle m$}}
\put(229,322){\makebox[1 cm][l]{$\scriptstyle n$}}
\put(214,195){\makebox[1 cm][l]{$\scriptstyle g$}}
\put(144,193){\makebox[1 cm][l]{$\scriptstyle d$}}
\put(140,232){\makebox[1 cm][l]{$\scriptstyle a$}}
\put(78,237){\makebox[1 cm][l]{$\scriptstyle e$}}
\put(76,273){\makebox[1 cm][l]{$\scriptstyle b$}}
\put(8,280){\makebox[1 cm][l]{$\scriptstyle f^\#$}}
\put(207,143){\makebox[1 cm][l]{$\scriptstyle c$}}
\put(258,128){\makebox[1 cm][l]{$\scriptstyle f$}}
\put(249,165){\makebox[1 cm][l]{$\scriptstyle c_1$}}
\put(318,120){\makebox[1 cm][l]{$\scriptstyle b_b$}}
\put(322,84){\makebox[1 cm][l]{$\scriptstyle e_b$}}
\put(382,76){\makebox[1 cm][l]{$\scriptstyle a_b$}}
\put(388,40){\makebox[1 cm][l]{$\scriptstyle d_b$}}
\put(424,17){\makebox[1 cm][l]{$\scriptstyle g_b$}}
\put(485,12){\makebox[1 cm][l]{$\scriptstyle b^*$}}
\end{picture}
\end{center}
\begin{center}
\begin{minipage}{12 cm}
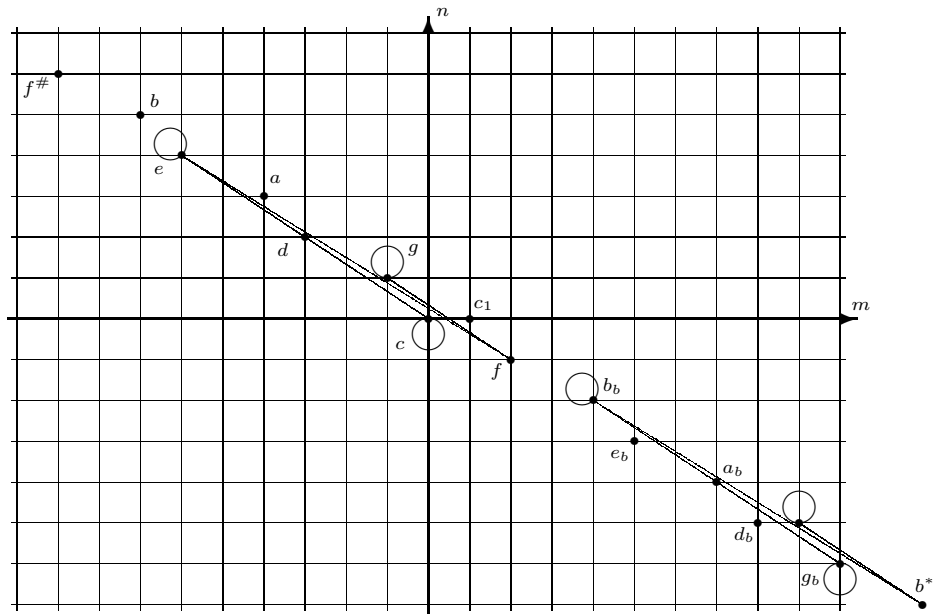
\mediumroman \baselineskip 9 pt Figure 5:  A transposition of the Beethoven theme. The harmonic space shows that the
note $\scriptstyle b$ changes in pitch through this transposition, we have called this new note $\scriptstyle b^*$
\end{minipage}
\end{center}
The transposition of the Beethoven theme such that it starts with $b_b$ uses the tones of the original
Pythagoras system as we have represented in Table \ref{tablebasictunes} and Table \ref{tablepythagorean}, except for the tone $b^*$. Indeed the note $b$
corresponds for this theme with the vector $(12, -7)$, which is not included in Table \ref{tablebasictunes} or Table \ref{tablepythagorean}. The tone $b$ as
included in Table \ref{tablebasictunes} or Table \ref{tablepythagorean} comes from the vector $(-7, 5)$, and not from the vector $(12, -7)$. Let us calculate the
difference between the tone coming from the vector $(-7, 5)$, denoted $b$, and the tone coming from vector $(12,
-7)$, denoted $b^*$. The difference is the length of the difference vector $(-19, 12) = (-7, 5) - (12, -7)$, and $\|(-19,
12)\| = (-19\log2 + 12\log3) \cdot 1200/\log2 = 23.460$ cent.

We can understand now very well what happens in general. Once a tone of a theme after transposition leaves the
rectangle of $10$ by $7$ (the rectangle that is shown in Figure \ref{tablepythagorean}) in which the traditional Phytagorean system is
constructed before being projected by the octave projection onto the octave, it will not coincide with the tones of
Table \ref{tablebasictunes} or Table \ref{tablepythagorean}. The difference will always be the same, namely 23.460 cent.

The musical pattern drawn in the vector space corresponding to the intrinsic harmonic spaces contains all the harmonic
content of a musical theme. Transposing just means transposing geometrically this pattern, as we have illustrated for
the Beethoven theme (see Figure 5). If this geometric pattern corresponding to a musical theme gets transposed over the
vector space, the projection into the octave, as illustrated in Figure 2, gives rise to different possible tones
corresponding to the same note, as it is the case for the note $b$ and the tones $b$ and $b^*$ in Figure 5.
Two such different tones are separated by a multiple of the Pythagorean COMMA. If a musical theme is represented in the
intrinsic harmonic space we always can find out easily, just watching and following the geometrical pattern, which notes
will correspond to which different tones in which parts of the musical theme. The rule is that we just have to minimize
the difference vectors in the vector space. This rule allows a uniquely determined and complete harmonic performance of
every possible musical theme, but equal notes will correspond to different tones for different parts of the theme as
prescribed by the geometry of the intrinsic harmonic space, and the way the musical theme is represented in this space.

\section{General Intrinsic Harmonic Spaces} \label{section04}
\noindent
In the foregoing section we have analyzed in detail the intrinsic harmonic space corresponding to the Pythagorean
system. We saw that this space is realized as a two dimensional vector space. Historically it is the Pythagorean system
that has given rise to the twelve half notes that exist in an octave in the tempered system, because $2^{19} \approx
3^{12}$. Our musical system is so much biased by this pattern of twelve half notes inside an octave, that is is
difficult for us to take fully into account the relative nature of this choice. That is also the reason that the
tempered system has become the system of reference now for all western music.

The story goes that the tempered system was propagated strongly by Johann Sebastian Bach. Indeed, one of the oevres of
Bach is called `Das Wohltemperierte Klavier' (The Well-Tempered Clavier), where he writes 24 fuga's and preludes for piano, each of them in a
different minor and major key. A closer look at the history
shows however that Bach was not really interested in equal tempering, as we know it now, but he was interested in a
system that would allow to play in all keys, but at the same time would conserve as much as possible the important
harmonies. At the time of Bach it was in effect meantone tuning that was used, and Bach wanted to show that if
fuga's are written well, it is possible to use this system for all keys. The complete equal temperament system was not
used in these times, because the consensus was that it sounded awful, out of tune and characterless. Only during the 19th
century, keyboard tuning drifted closer and closer to equal temperament over the
protest of many of the more sensitive musicians. Only in 1917 was a method devised for tuning exact equal temperament.

\subsection{Meantone Tuning}
\noindent
Meantone tuning can be considered to be Europe's most successful tuning. It appeared sometime around
the late 15th century, and was used widely through the early 18th century. It survived in the tuning of English organs,
all the way through the 19th century.

The principle of meantone tuning is that preserving the consonance of the thirds ($c$ to $e$, $f$ to $a$, $g$ to $b$) is more
important than preserving the purity of the fifths ($c$ to $g$, $f$ to $c$, $g$ to $d$). The acoustical
reason for this preference is that the notes in a slightly out-of-tune third, being closer together than those in a fifth,
create faster and more disturbing beats than those in a slightly out-of-tune fifth \cite{rayleigh}. In a perfect harmonic third,
the two strings vibrate at a frequency ratio of 5 to 4. In section \ref{sec:harmonicpossibilities} we analyzed already the
possibility of the perfect harmonic third (see Table \ref{tableaddingthird}), and found that its value in cent equals 386.314. As we mentioned
already in section
\ref{sec:harmonicpossibilities}, the fact that three perfect fifths do not give an octave, since $3 \times 386.314 = 1158.941 <
1200$, was one of the basic problems of perfect harmony that musicians have been confronted with. This means indeed that it is
not possible to adopt a tuning system where, for example, the three thirds $c$ to $e$, $e$ to $a_b$, and $a_b$ to $c_1$, would
all be tuned as perfect thirds. Meantone tuning tried to find a solution where as much as possible of the thirds are tuned as
perfect thirds, by giving up somewhat on the perfect tuning of the firths as compared to Pythagorean tuning.

There is not one unique meantone tuning system, several variations on the basic ideas have been tried out. Let us explain the
one invented by Pietro Aaron in 1523, as counted in \cite{owen} (see Table \ref{tablemeantone} for a presentation of this meantone tuning
system).

\smallskip
\noindent
Fixing $c_1, e$ and $a_b$:

\smallskip
\noindent
We start with the octave $c_1/c =2$, hence $c$ and $c_1$ are separated by 12.000 cent, as usual. The second step is to
tune a perfect third, which means that $e/c = 5/4$, and hence $c$ and $e$ are separated by 386.314 cent. Also the third with $e$
as basic note is tuned in perfect harmony, which means that $a_b/e = 5/4$ and hence $a_b/c = a_b/e \cdot e/c = 25/16$. This
means that $c$ and $a_b$ are separated by $2 \times 386.314 = 772.627$ cent. There cannot be a perfect third with $a_b$ as basic
note, if we want to respect the octave, hence the $a_b$ key is sacrificed, it will not be used in the meantone tuning system.
More specifically $c_1/a_b = c_1/c \cdot c/a_b = 2 \cdot 16/25 = 32/25$. This means that $a_b$ and $c$ are separated by 427.372
cent, which is way too much for a third. Since this distance comes however from a fraction that is still simple, namely 32/25, it
sounds all right, but not like a third. This is the simple part of the meantone tuning. 

\smallskip
\noindent
Fixing $d, f^\#$ and $b_b$:

\smallskip
\noindent
Remark that in the Pythagorean system we
have $d/c = 9/8$, and we also have $c/b_b = 9/8$. This means that $d/b_b = 81/64$. This means that $b_b$ and $d$ are separated
in the Pythagorean system by 407.820 cent, and hence do not form a perfect third, because then they would have to be separated
by 386.314 cent. To rescue the perfect third $b_b$ to $d$ it is decided to sacrifice the Pythagorean second of $9/8$. More
specifically, it is decided to tune the seconds by just taking half of the perfect third. This does not correspond any longer to
a fraction, which means that in meantone tuning for the first time there is this departure from fractions, as later happens
completely for tempered tuning. Hence the note $d$ is tuned with distance 386.314/2 = 193.157 cent from $c$ and distance
386.314/2 = 193.157 cent from $e$. It is this decision that has given its name to the meantone tuning, taking the meantone for
$d$ of $e$ and $c$. In a similar way $f^\#$ is put in between $e$ and $a_b$. This gives a difference between $c$ and $f^\#$ of
386.314 + 193.157 = 579.471 cent. By doing so we have obviously also created a perfect third from $d$ to $f^\#$. We have no
intention to create a harmonious $a_b$ key, because the distance between $a_b$ and $c_1$ is already spoiled by our compromise
between the perfect thirds and the octave. This means that we can as well choose $b_b$ in such a way that also the $b_b$ key
contains a perfect third. This means, for example, that we have to choose the distance between $b_{b-1}$ and $c$ equal to
193.157 cent, which means that the distance between $c$ and $b_b$ must equal 1200.000 - 193.157 = 1006.843 cent. 

The meantone
tuning wants to pay attention also to the minor thirds, as well to the fifths of course. A perfect minor third corresponds to the
ratio 6/5, which is a distance of 315.641 cent. A perfect fifth corresponds to the ratio 3/2, which is a distance of
701.955. Some harmony theory comes in here. For the overall harmony within one specific key there are two minor thirds of other
keys that are the most important ones. For example for the overall harmony in key $c$, the most important minor thirds are the
one of key $a$, called the relative one, and the one of key $e$. A perfect minor third of key $e$ would fix $g$ at
cent distance 386.314 + 315.641 = 701.955 cent, which would create perfect harmony within key $c$, because also the fifth is
perfect in this case. This is however not possible for other keys, and a compromise has to be made if also in other keys one
wants to keep close to the perfect minor thirds and the perfect fifths.

\smallskip
\noindent
Fixing $g$:

\smallskip
\noindent
The cent distance between $e$ and $g$ is chosen to
be 310.500 cent, which fixes $g$ at 386.314 + 310.500 = 696.814 cent. This creates two fifths, one in key $c$ with value 696.814
cent, and another in key $g$ with value 1200.000 + 193.157 - 696.814 = 696.343 cent. Both fifths are too small, but sufficiently
close to the perfect fifth to be all right. 

\smallskip
\noindent
Fixing $a$:

\smallskip
\noindent
The cent distance between $a$ and $c_1$ is chosen to be 310.300
cent, which fixes $a$ at a distance 1200.000 - 310.300 = 889.700 cent, and creates again two fifths. One in key $d$ of value
889.700 - 193.157 = 696.543 cent and another in key $a$ with value 1200.000 + 386.314 - 889.700 = 696.614 cent. Fixing $a$ we
have also fixed the minor third in key $f^\#$, with value 889.700 - 579.471 = 310.229 cent, which is an important minor key for
the overall harmony in key $d$. 

\smallskip
\noindent
Fixing $b$:

\smallskip
\noindent
We choose $b$ by giving the value of 310.300 cent to the minor third in $b$, which is the
relative minor third to the third in key $d$, and hence the most important minor third for the overall harmony in key $d$. This
fixes $b$ at 193.157 - 310.300 + 1200.000 = 1082.857 cent. The fixing of $b$ creates two new fifths, namely the fifth of key
$b$, with value 579.471 - 1082.857 + 1200.000 = 696.614 cent, and the fifth of key $e$, with value 1082.857 - 386.314 = 696.543
cent. Both fifths are reasonable approximations of the perfect fifth. The fixing of $b$ also introduces a new minor third in key
$a_b$ with value 1082.857 - 772.627 = 310.230 cent.

\smallskip
\noindent
Fixing $e_b$:

\smallskip
\noindent
In a similar way the distance of $e_b$ is chosen such that the minor
third in key
$c$, which is the relative minor third of the third in key $e_b$ fits the value 310.300. Hence this fixes
$e_b$ on 310.300 cent. This creates again two new fifths, the one in key $e_b$, with value 1006.843 - 310.300 = 696.543 cent,
and the one in key $a_b$, with value 310.300 + 1200.000 - 772.627 = 737.673 cent. And here we have arrived at a point where the
meantone tuning system fails completely. Key $a_b$ has a third and a fifth that are both way to big. It will not be possible to use
this key within the meantone tuning system. The fixing of $e_b$ also creates a new minor third, namely the one of key $e_b$
itself, with value 579.471 - 310.300 = 269.171 cent. This is also a bad value.

\begin{table}[h]
\caption{Meantone tuning and equal temperament tuning compared}
\begin{center}
\begin{tabular}{lllllll} \hline 
\multicolumn{3}{l}{} & \multicolumn{2}{l}{\bf Meantone} & \multicolumn{2}{l}{\bf Equal Temperament} \\ \hline
&{\it Tone Ratio} & {\it Example} &  {\it Freq. Fraction} & {\it Cent} & {\it Freq. Fraction} & {\it Cent} \\ \hline
0 & Basic & $c/c$ & 1 & 0.000 & 1 & 0.000 \\ 
1 & & $d_b/c$ & no fraction & 76.014  & $\root 12 \of {2}$ = 1.0594 & 100.00 \\ 
2 & Second & $d/c$ & no fraction & 193.157 & $(\root 12 \of {2})^2$ = 1.1224 & 200.00\\ 
3 & & $e_b/c$ & no fraction & 310.300 &$(\root 12 \of {2})^3$ = 1.1892 & 300.00\\ 
4 & Third & $e/c$ & 5/4 = 1.25 & 386.314 & $(\root 12 \of {2})^4$ = 1.2599 & 400.00\\ 
5 & Fourth & $f/c$ & 4/3 = 1.3333 & 503.4 & $(\root 12 \of {2})^5$ = 1.3348 & 500.00 \\ 
6 & & $f^\#/c$ & no fraction & 579.471 & $(\root 12 \of {2})^6$ = 1.4142 & 600.00 \\ 
7 & Fifth & $g/c$ & no fraction & 696.800 & $(\root 12 \of {2})^7$ = 1.4983 & 700.00 \\ 
8 & & $a_b/c$ & 25/16 = 1.5625 & 772.627 & $(\root 12 \of {2})^8$ = 1.5874 & 800.00 \\ 
9 & Sixth & $a/c$ & no fraction & 889.700 & $(\root 12 \of {2})^9$ = 1.6817 & 900.00\\ 
10 & & $b_b/c$ & no fraction & 1006.843 & $(\root 12 \of {2})^{10}$ = 1.7817 & 1000.00\\ 
11 & Seventh & $b/c$ & no fraction =  & 1082.857 & $(\root 12 \of {2})^{11}$ = 1.8877 & 1100.00\\ 
12 & Octave & $c_1/c$ & 2 & 1200.00 & 2 & 1200.00 \\ \hline
\end{tabular}
\label{tablemeantone}
\end{center}
\end{table}

\subsection{The Intrinsic Harmonic Space I(2, 5)}

\noindent
The intrinsic harmonic space related to the Pythagorean tuning system only uses the prime numbers 2 and 3, and that is
why we will denote it $I(2, 3)$. Meantone tuning introduces the perfect third by allowing the prime number 5 to play a role in the harmony. Let us see how the perfect harmonic space I(2, 5) would look like if built along the same principles that the one we used to built the intrinsic harmonic space I(2, 3). Within I(2, 3) is is the fact that $2^{19} \approx 2^{12}$ that gives rise the tuning within a system of 12 tones. Hence lets look for a power of 5 that is close to a power of 2. In fact, this happens more quickly than in the case of the I(2, 3) system. Indeed $5^3 = 125$ and $2^7 =128$. Let us make a construction which is the equivalent with respect to the space I(2, 5) of what the Pythagorean tuning is with respect to the space I(2, 3). We take $c$ as basic tone. The first tone of this construction is the perfect third $e$ corresponding to the point (-2, 1) of the vector space corresponding to I(2, 5) and hence with ratio $5/4$ with respect to $c$. The second tone corresponding to the point (3, -1), and hence with ratio $8/5$ and hence 813.686 cent, situates itself close to the Pythagorean $a_b$ with 792.180 cent, which hence gives a difference of 21.506 cent. Let us denote this tone $a_b(5)$, the reference to 5 indicating that we are working now in I(2, 5), and this also means that we can indicate the perfect $e$ by $e(5)$. Let us see now we are confronted with a very similar type of trouble than the one we encountered already with the attempt of putting the Pythagorean intrinsic harmonic space I(2, 3) onto the one dimensional octave keyboard. Indeed, the third tone that gives rise to a projection inside the octave is given by the point (-4, 2), hence the ratio 25/16, which is 772.627 cent. This is again in the region of the tempered $a_b$ at 800.000 cent, which means that we have two candidates now for a $a_b$ within the octave projection of I(2, 5), one at 813.686 cent, which we called $a_b(5)$ and another one at 772.627, which we will call $a_b^*(5)$. The difference between both, 813.686 - 772.627 = 41.059 cent, is the value of the $5-COMMA$, which is the comma of the 5-system in analogy with the Pythagorean comma. The fourth tone corresponds to the point (5, -2), hence the ratio 32/25 which is 427.373 cent and is a second perfect version of the tempered $e$ at 400.000 cent, which we will call $e^*$. Its difference with the perfect $e$ equals 427.373 - 386.314 = 41.059 cent, again the value of the $5-COMMA$. 

This means that only three tones can be constructed harmonically within the construction which is the equivalent one with respect to I(2, 3) as the Pythagorean is with respect to I(2, 3), which we could expect in fact, because the power of 5 that comes close to a power of 2 is 3. The tone system that results in this way is too small to be able to lead to a tone system that can be used to play melodies. However the meantone tuning system uses some of the tones that result in this way. However, meantone tuning also makes use of the ratio $6/5$ for the perfect minor third. Probably nobody was aware of this at that time, but the use of a perfect minor third was a first step towards the use of a higher than two dimensional harmonic space. We will analyse this space in the next section.

\subsection{The Intrinsic Harmonic Space I(2, 3, 5)}
\noindent
To introduce the harmony of the perfect third and the perfect minor third within the intrinsic harmonic spaces we have to
consider the harmonic space built by using the prime numbers 2, 3 and 5. This is a three dimensional space, each vector of it is
written as a triple $(x, y, z)$ of three numbers. Meanwhile it is clear that music written in this new intrinsic harmonic space should make use directly of the vectors and the difference vectors within this three dimensional vector space. However, because of the ingrained tradition of the one dimensional octave system corresponding to traditional keyboards we are in need to look at the octave projection of this new intrinsic harmonic space I(2, 3, 5) to be able to see and understand its reach. Hence, let us mention some of the octave projections of I(2, 3, 5) 
with $c$ as basic key. Hence in this case $c$ is represented by the vector $(0, 0, 0)$. The vector $(1, 0, 0)$ represents $c_1$, and the vector $(-1, 1, 0)$ represents $g$. The vector $(-2, 0, 1)$
represents $e$ when tuned as a perfect third, while the vector $(-6, 4, 0)$ represents $e$ when tuned within the Pythagorean
system. We mentioned already that the fact that we have twelve half notes in a octave is related to the fact that $2^{19} \approx
3^{12}$. The vector $(-19, 12, 0)$ corresponds to the new $c$, denoted $c^*$, within the Pythagorean tuning. The
difference between $(0, 0, 0)$ and $(-19, 12, 0)$ in the octave projection equals 23.460 cent and is called the Pythagorean $COMMA$. We also mentioned that the I(2, 5) intrinsic harmonic space attributes 2 extra tones to each note, one being the perfect third of each of the considered notes and the other being what is the $a_b(5)$ with respect to $c$ taken with respect to each of the considered notes. However, meantone tuning also had introduced the perfect minor third, at a fraction of 6/5 with respect to the $c$. This tone corresponds to the vector (1, 1, -1), which is a genuine vector of the intrinsic harmonic space I(2, 3, 5), not existing in neither I(2, 3) and nor I(2, 5).

In Table \ref{I(2,3,5)} we collect some of the tones that result if anyhow such a projection is attempted, and we take $c$ as a basic tone. Hence we have added the perfect thirds and perfect minor thirds for some of the Pythagorean tones, more specifically for $c$, $d$, $e$, $f$, $g$, $a$ and $b$. It is interesting to note that some of these `new' tones come closer to the tempered ones than the original Pythagorean tones and the ones of the meantone system. For example, the tone (-7, 3, 1) is 7.821 cent away from the tempered $d_b$, while the Pythagorean $d_b^{pyth}$ is 9.775 cent away from the tempered, while the meantone $d_b^{mean}$ is 23.986 cent away from the tempered. The tone (-10, 5, 1) is 3.911 cent away from the tempered $e_b$, while the Pythagorean and the meantone are respectively 5.865 and 15.641 cent away. The Pythagorean Fourth and Fifth remain unchallenged and are closer to the tempered $f$ and $g$ than any of the new tones. However, the tone (-5, 2, 1) is 9.776 cent away from the tempered $g_b$, while the Pythagorean and tempered values differ  by 11.730. The tone (-8, 4, 1) is 5.866 cent away from the tempered $a_b$, while the Pythagorean $a_b^{pyth}$ is 7.820 cent away.

\begin{longtable}{p{1.7cm}p{4.7cm}p{1.7cm}p{2.7cm}p{1.7cm}}
\caption{The intrinsic harmonic space I(2, 3, 5)} \\ \hline
{\it Vector} & {\it Tone Ratio} & {\it Example} & {\it Projection Value} & {\it Deviation} \\ \hline
 (0, 0, 0) & {\small Basic} & $c$ & 0.000 & 0.000 \\ 
(-4, 4, -1) &Meantone Minor Third for $a$ & & 21.506 & 21.506 \\ \hline
 (8, -5, 0) & & $d_b^{pyth}$ & 90.225 & 9.775 \\ 
(-7, 3, 1) & Meantone Third for $a$ & & 92.179 & 7.821 \\  \hline
 (-3, 2, 0) & Pythagorean Second & $d^{pyth}$ & 203.910 & 3.910 \\
 (-7, 6, -1) & Meantone Minor Third for $b$ & & 225.416 & 25.416 \\ \hline
 (5, -3, 0) & Pythagorean Minor Third & $e_b^{pyth}$ & 294.135 & 5.865\\ 
(-10, 5, 1) & Meantone Third for $b$ & & 296.089 & 3.911 \\ 
(1, 1, -1) & Meantone Minor Third for $c$ & $e_b^{mean}$ & 315.641  & 15.641\\ \hline
(-2, 0, 1) & Meantone Third for $c$ & $e^{mean}$ & 386.314 & 13.686 \\ 
(-6, 4, 0) & Pythagorean. Third & $e^{pyth}$ & 407.820 & 7.820 \\ \hline
 (2, -1, 0) & Pythagorean Fourth & $f^{pyth}$ & 498.045 & 1.955 \\
 (-2, 3, -1) & Meantone Minor Third for $d$ & & 519.551 & 19.551 \\ \hline
(-5, 2, 1) & Meantone Third for $d$ & & 590.224 & 9.776 \\ 
 (-9, 6, 0) & & $f^{\#pyth}$ & 611.730 & 11.730 \\  \hline
(-1, 1, 0) & Pythagorean Fifth & $g^{pyth}$ & 701.955 & 1.955 \\
 (-5, 5, -1) &Meantone Minor Third for $e$  & & 723.461 & 23.461 \\  \hline 
(7, -4, 0) & & $a_b^{pyth}$ & 792.180  & 7.820 \\ 
(-8, 4, 1) & Meantone Third for $e$ & & 794.134 & 5.866 \\ 
(3, 0, -1) & Meantone Minor Third for $f$ & & 813.686 & 13.686 \\  \hline
(0, -1, 1) & Meantone Third for $f$ & $a^{mean}$ & 884.359 & 15.641 \\ 
(-4, 3, 0) & Pythagoran Sixth  & $a^{pyth}$ & 905.865 & 5.865 \\  \hline
 (4, -2, 0) & & $b_b^{pyth}$ & 996.090 & 3.910 \\ 
 (0, 2, -1) & Meantone Minor Third for $g$ & & 1017.596 & 17.596 \\  \hline
(-3, 1, 1) & Meantone Third for $g$ & & 1088.269 & 11.731 \\ 
 (-7, 5, 0) & Pythagorean Seventh & $b^{pyth}$ & 1109.775 & 9.775 \\  \hline 
 (1, 0, 0) & {\small Octave} & $c_1$ & 1200.000 & 0.000 \\ \hline
 \label{I(2,3,5)}
 \end{longtable}

\subsection{The Beethoven Theme in I(2, 3, 5)}
\noindent
Let us write a version of the Beethoven theme in the intrinsic harmonic space I(2, 3, 5), making use of the presence of the perfect third which is absent in the Pythagorean intrinsic harmonic space I(2, 3).

\begin{table}[h]
\caption{The Beethoven theme in I(2, 3, 5). The first column represents the vectors, the second column the naming of the tones, the third column the difference vectors, and the fourth column the
actions as named in the traditional way}
\begin{center}
\begin{tabular}{lllll} \hline
Vector & Tone & Difference Vector & Action \\ \hline
 (-2, 0, 1) & $e$ & (4, -1, -1) = (2, -1, 0) - (-2, 0, 1)  & half tone up \\ \hline
 (2, -1, 0) & $f$ & (-3, 2, 0) = (-1, 1, 0) - (2, -1, 0)& tone up \\ 
 (-1, 1, 0) & $g$ & (0, 0, 0) = (-1, 1, 0) - (-1, 1, 0) & stay \\ 
 (-1, 1, 0) &  $g$ & (3, -2, 0) = (2, -1, 0) - (-1, 1, 0) & tone down \\ 
 (2, -1, 0) & $f$ & (-4, 1, 1) =  (-2, 0, 1) - (2, -1, 0)& half tone down \\ 
 (-2, 0, 1) & $e$ & (-1, 2, -1) = (-3, 2, 0) - (-2, 0, 1) & tone down \\ 
 (-3, 2, 0) & $d$ & (3, -2, 0) = (0, 0, 0) - (-3, 2, 0)& tone down \\ 
 (0, 0, 0) & $c$ & (0, 0, 0) = (0, 0, 0) - (0, 0, 0) & stay \\ 
 (0, 0, 0) & $c$ & (-3, 2, 0) = (-3, 2, 0) - (0, 0, 0) & tone up \\ 
(-3, 2, 0) & $d$ & (1, -2, 1) = (-2, 0, 1) - (-3, 2, 0)  & tone up \\ 
 (-2, 0, 1) & $e$ & (0, 0, 0) = (-2, 0, 1) - (-2, 0, 1)& stay \\ 
 (-2, 0, 1) & $e$ & (-1, 2, -1) = (-3, 2, 0) - (-2, 0, 1)& tone down \\ 
 (-3, 2, 0) & $d$ & (0, 0, 0) = (-3, 2, 0) - (-3, 2, 0)& stay \\ 
 (-3, 2, 0) &  $d$ &  &  \\ \hline
\end{tabular}
\label{tablebeethovenmeantone}
\label{}
\end{center}
\end{table}

The difference with the theme in the Pythagorean tone system can be seen very easily now and i represented in Table \ref{tablebeethovenmeantone}. We have used the perfect third for $e$ instead of the Pythagorean third. The other tones have remained Pythagorean, because for them the Pythagorean tones are the best choice. Remark that the actions `half tone up' and `half tone down' are characterized by different vectors than this was the case for the Beethoven tune in the Pythagorean space I(2, 3). Also two different actions correspond to `tone up' represented by the vectors (-3, 2, 0) and (1, -2, 1), and the opposite actions (3, -2, 0) and (-1, 2, -1) correspond to `tone down'.

Again, like in the Pythagorean case, this Beethoven theme can be transposed to any starting tone within the intrinsic harmonic space I(2, 3, 5), and keep its perfect harmonic content.

\section{The Prime Number 7 and the Space I(2, 3, 5, 7)}
\noindent
Pythagorean tuning only considers the prime numbers 2 and 3. Meantone tuning had discovered the prime number 5 for some of its tones. What if we allow the prime number 7 to join the scene? In principle, this should give rise to harmonies that are of a completely new type, neither present within the Pythagorean system nor within the meantone system. 

\subsection{The Space I(2, 7)}
\noindent
Let us start with the most simple of all ways to incorporate the prime number 7, namely in the intrinsic harmonic space I(2, 7). The question we have to investigate first is whether there is a power of 2 that comes close to a power of 7. The space I(2, 3) gives rise to 12 tones because $2^{19} \approx 3^{12}$. The space I(2, 5) gives rise to 3 tones because $2^7 \approx 5^3$. For I(2, 7) we get $2^{73} \approx 7^{26}$. This means that if we construct a set of tones in a similar way than is done within the Pythagorean system, we will find 26 tones in an octave for I(2, 7). See Table \ref{table:I(2,7)} for the construction of the 26 tones.

\begin{longtable}{p{2cm}p{2.7cm}p{1.7cm}p{2.7cm}p{1.7cm}}
\caption{The intrinsic harmonic space I(2, 7)} \\
{\it Vector} & {\it Tone Ratio} & {\it Example} & {\it Projection Value} & {\it Deviation} \\ \hline
(0, 0, 0, 0) & Basic & $c$ & 0.000 & 0.000 \\ 
(-14, 0, 0, 5) & &  & 44.130 & 44.130 \\ \hline
(-28, 0, 0, 10) & & & 88.259 & 11.741 \\
(31, 0, 0, -11) & & & 142.915 & 42.915 \\ \hline
(17, 0, 0, -6) & {\small Second} &  $d$ & 187.045 & 12.955 \\ 
(3, 0, 0, -1) & & & 231.174 & 31.174 \\ \hline
(-11, 0, 0, 4) & & & 275.304 & 24.696 \\ 
(-25, 0, 0, 9) & {\small Minor Third} & $e_b$ & 319.433 & 19.433 \\ \hline
(34, 0, 0, -12) & & & 374.089 & 25.911 \\ 
(20, 0, 0, -7) & {\small Third} & $e$ & 418.218 & 18.218 \\ \hline
(6, 0, 0, -2) & & & 462.348 & 37.652 \\ 
(-8, 0, 0, 3) & {\small Fourth} & $f$ & 506.478 & 6.478 \\ \hline
(-22, 0, 0 8) & & & 550.607 & 49.393 \\
(-36, 0, 0, 13) & & $g_b$ & 594.737 & 5.263 \\ 
(37, 0, 0, -13) & & $f^\#$ & 605.263 & 5.263 \\
(23, 0, 0, -8) & & & 649.393 & 49.393 \\ \hline
(9, 0, 0, -3) & {\small Fifth} & $g$ & 693.522 & 6.478 \\  
(-5, 0, 0, 2) & & & 737.652 & 37.652 \\ \hline
(-19, 0, 0, 7) & & $a_b$ & 781.781 & 18.209 \\ 
(-33, 0, 0, 12) & & & 825.911 & 25.911 \\ \hline
(26, 0, 0, -9) & {\small Sixth} & $a$ & 880.567 & 19.433 \\ 
(12, 0, 0, -4) & & & 924.696 & 24.696 \\ \hline
(-2, 0, 0, 1) & & & 968.826 & 31.174 \\ 
(-16, 0, 0, 6) & & $b_b$ & 1012.956 & 12.956 \\ \hline
(-30, 0, 0, 11) & & & 1057.085 & 42.915 \\
(29, 0, 0, -10) & {\small Seventh} & $b$ & 1111.741 & 11.741 \\ \hline
(15, 0, 0, -5) & & & 1155.870 & 44.130 \\ 
(1, 0, 0, 0) & {\small Octave} & $c_1$ & 1200 & 0 \\ \hline
\label{table:I(2,7)}
\end{longtable}
\noindent
We can construct the intrinsic harmonic spaces I(2, 3, 7), I(2, 5, 7) which will be three dimensional vector spaces each of them, and also the intrinsic harmonic space I(2, 3, 5, 7) which is a four dimensional vector space, combining all the constructions we did before and using the same method. Already the space I(2, 3, 5) will allow to play a type of harmony that never has been able to be realized before, and was attempted to by the different versions of the meantone tuning. However, when we also introduce the harmonies offered by the intrinsic harmonic space I(2, 3, 5, 7) a totally new type of harmony becomes available able to produce chord and tones that never before were able to be produced in a systematic way.

\subsection{Musical instruments for perfect harmonic spaces}
\noindent
The reason that the old problem of perfect tuning was never solved is strictly related to the limitation that sets in if one wants to achieve such tuning using a musical instrument that generates tones with a keyboard that is one dimensional. In principle, music played on violins produce such harmonies. However, violin music is also limited in performance and especially in notation by the 12 notes of the octave. Although a violin can play all tones in principle there is no systematic way to indicate these tones let alone to form chords and keys with them. All of this becomes possible with the help of the intrinsic harmonic spaces presented here. If, in a first step, we limit ourselves to the intrinsic harmonic space I(2, 3) built around the Pythagorean scale, we can better understand what the real limitation is. In Figure 2 it becomes clear that classical thinking takes place in a one dimensional projection of the true harmonic space, which is two dimensional. So it is the attempt to squeeze harmony into the linear one-dimensional space of frequencies, the space that goes from low frequencies linear on a keyboard to high frequencies, on the keyboard from left to right, that makes it impossible. The intrinsic harmonic space I(2, 3) needs two dimensions. If we build a musical instrument that is two dimensional instead of one dimensional we can play the total intrinsic harmonic space I(2, 3) on that musical instrument. What does such a musical instrument look like? Consider again Figure 2 where the plane is shown in which I(2, 3) lies. Imagine that each point on this plane with integer coordinates, i.e., where the lines intersect, is a key such that when this key is pushed the tone corresponding to the point is played. Playing these keys then allows us to directly play the intrinsic harmonic space I(2, 3). The Beethoven theme worked out in Figure 3, Figure 4 and Figure 5 gives a good example of how to play such an instrument. Note that the tones are not ordered from low to high frequency, but the order is according to the I(2, 3) harmony.

Note, however, that for I(2, 3, 5) we already need three dimensions. This could still be realized with a physical keyboard-like instrument since the 5 harmony only adds a few notes. For example, one could work with a pedal that switches for the vertical up and down of the dimension where the 5 harmony belongs. And the 2, 3 harmony is then realized in the plane with keys at the points of the plane, as in Figure 2. However, if we go to higher intrinsic harmonic spaces, such as I(2, 3, 5, 7), then four dimensions are required and a physical instrument becomes difficult to realize. However, now that we have computers available, even this limitation can be removed. Programming tones by using the vectors of the vector spaces as denoting a tone is no problem at all in a computer, even for I(2, 3, 4, 7). Even I(2, 3, 4, 7, 11) can be explored that way. In further work, we plan to program some well-known pieces of music in such a way, in order to get to performances of these pieces of music with perfect harmony, which was probably never heard before by human ears.

\end{document}